\documentclass[letterpaper, 12pt]{article}[2000/05/19]
\usepackage[english]{babel}
\usepackage{amsfonts,amsmath,amssymb,amsthm,latexsym,amscd,mathrsfs}
\usepackage{ifthen,cite}
\usepackage[bookmarksnumbered=true]{hyperref}
\hypersetup{pdfpagetransition={Split}}

\newcommand{\evenhead}{Author \ name}
\newcommand{\oddhead}{Article \ name}
\newcommand{\theArticleName}{Article \ name}

\newcommand{\FirstPageHeading}[1]{\thispagestyle{empty}%
\noindent\raisebox{0pt}[0pt][0pt]{\makebox[\textwidth]{\protect\footnotesize \sf }}\par}

\newcommand{\ArticleName}[1]{\renewcommand{\theArticleName}{#1}\vspace{-2mm}\par\noindent {\LARGE\bf  #1\par}}
\newcommand{\Author}[1]{\vspace{5mm}\par\noindent {\Large  #1\par} \par\vspace{2mm}\par}
\newcommand{\Address}[1]{\vspace{2mm}\par\noindent {\it #1} \par}
\newcommand{\Email}[1]{\ifthenelse{\equal{#1}{}}{}{\par\noindent {\rm E-mail: }{\it  #1} \par}}
\newcommand{\URLaddress}[1]{\ifthenelse{\equal{#1}{}}{}{\par\noindent {\rm URL: }{\tt  #1} \par}}
\newcommand{\EmailD}[1]{\ifthenelse{\equal{#1}{}}{}{\par\noindent {$\phantom{\dag}$~\rm E-mail: }{\it  #1} \par}}
\newcommand{\URLaddressD}[1]{\ifthenelse{\equal{#1}{}}{}{\par\noindent {$\phantom{\dag}$~\rm URL: }{\tt  #1} \par}}

\newcommand{\Abstract}[1]{\vspace{6mm}\par\noindent\hspace*{10mm}
\parbox{140mm}{\small {\bf Abstract.} #1}\par}
\newcommand{\Keywords}[1]{\vspace{3mm}\par\noindent\hspace*{10mm}
\parbox{140mm}{\small {\bf Key words:} \rm #1}\par}
\newcommand{\Classification}[1]{\vspace{3mm}\par\noindent\hspace*{10mm}
\parbox{140mm}{\small {\it 2010 Mathematics Subject Classification:} \rm #1}\vspace{3mm}\par}
\newcommand{\ShortArticleName}[1]{\renewcommand{\oddhead}{#1}}
\newcommand{\AuthorNameForHeading}[1]{\renewcommand{\evenhead}{#1}}

\long\def\@makecaption#1#2{
  \sbox\@tempboxa{\small \textbf{#1.}\ \ #2}%
  \ifdim \wd\@tempboxa >\hsize
    {\small \textbf{#1.}\ \ #2}\par \else
    \global \@minipagefalse
    \hb@xt@\hsize{\hfil\box\@tempboxa\hfil}%
  \fi \vskip\belowcaptionskip}

\def\numberwithin#1#2{\@ifundefined{c@#1}{\@nocounterr{#1}}{%
  \@ifundefined{c@#2}{\@nocnterr{#2}}{%
  \@addtoreset{#1}{#2}%
  \toks@\@xp\@xp\@xp{\csname the#1\endcsname}%
  \@xp\xdef\csname the#1\endcsname
    {\@xp\@nx\csname the#2\endcsname.\the\toks@}}}}
\def\E^#1{{\buildrel #1 \over\vee}}

{\theoremstyle{definition}

}

\usepackage{tikz}

\begin{document}

\FirstPageHeading{V. I. Gerasimenko, I. V. Gapyak}

\ShortArticleName{Propagation of correlations}

\AuthorNameForHeading{V. I. Gerasimenko, I. V. Gapyak}


\ArticleName{\textcolor{blue!50!black}{Propagation processes of correlations\\ of hard spheres}}

\Author{V. I. Gerasimenko\footnote{E-mail: \emph{gerasym@imath.kiev.ua}}}

\Address{\hspace*{1mm} Institute of Mathematics of the NAS of Ukra\"{\i}ne,\\
    \hspace*{2mm}3, Tereshchenkivs'ka Str.,\\
    \hspace*{2mm}01004, Ky\"{\i}v, Ukra\"{\i}ne}

\Author{I. V. Gapyak\footnote{E-mail: \emph{gapjak@ukr.net}}}

\Address{\hspace*{2mm} Taras Shevchenko National University of Ky\"{\i}v,\\
               \hspace*{3mm}Department of Mathematics and Mechanics,\\
               \hspace*{2mm} 4, Academician Glushkov ave.,\\
               \hspace*{2mm} 03127, Ky\"{\i}v, Ukra\"{\i}ne}

\bigskip

\Abstract{
The paper develops an approach to the description of the evolution of correlations for many hard spheres
based on a hierarchy of evolution equations for the cumulants of the probability distribution function
governed by the Liouville equation. It is established that the constructed dynamics of correlations underlies
the description of the evolution of the states of many hard spheres described by the BBGKY hierarchy for reduced
distribution functions or the hierarchy of nonlinear evolution equations for reduced correlation functions.
As an application of the developed approach to describing the evolution of the state of many hard spheres within
the framework of dynamics of correlations, the challenges of the derivation of kinetic equations are discussed.
}

\bigskip
\vspace{2pc}
\Classification{82C22, 82C40, 35Q20.}

\makeatletter
\renewcommand{\@evenhead}{
\hspace*{-3pt}\raisebox{-15pt}[\headheight][0pt]{\vbox{\hbox to \textwidth {\thepage \hfil \evenhead}\vskip4pt \hrule}}}
\renewcommand{\@oddhead}{
\hspace*{-3pt}\raisebox{-15pt}[\headheight][0pt]{\vbox{\hbox to \textwidth {\oddhead \hfil \thepage}\vskip4pt\hrule}}}
\renewcommand{\@evenfoot}{}
\renewcommand{\@oddfoot}{}
\makeatother

\newpage
\vphantom{math}

\protect\textcolor{blue!50!black}{\tableofcontents}
\textcolor{blue!50!black}{\section{Introduction}}

Recently, mainly in connection with the problem of the rigorous derivation of kinetic equations from
the underlying particle dynamics \cite{CGP97}-\cite{SR12}, a number of papers \cite{TSRS}-\cite{GG18}
have appeared discussing possible approaches to describing the evolution of the state of many-particle
systems, in particular, many hard spheres.

As is known \cite{CGP97}-\cite{Sp91}, the evolution of the state of finitely many hard spheres is
traditionally described by a probability distribution function governed by the Liouville equation.
This article is developed an alternative approach to the description of the evolution of the state,
which consists of the use of functions defined as cumulants of the mentioned probability distribution
function. The cumulants of the probability distribution function are interpreted as correlations of
the states of hard spheres, and further, the term correlation functions are used for them. The evolution
of the correlation functions is governed by the Liouville hierarchy for hard spheres constructed in
the article.

One more approach that allows describing the evolution of states of the systems both from a finite and
from an infinite number of particles \cite{CGP97}-\cite{Sp91} is to describe the state by means of a
sequence of so-called reduced distribution functions (marginals \cite{SR12}-\cite{DS}) governed by the
BBGKY (Bogolyubov--Born--Green--Kirkwood--Yvon) hierarchy \cite{CGP97}-\cite{SR12}. Note that this
method is traditionally used in the rigorous derivation of the Boltzmann kinetic equation from the
dynamics of hard spheres \cite{G58}-\cite{PG90}. An alternative of the approach to such a description
of a state is based on a sequence of functions determined by the cluster expansions of reduced distribution
functions that are, by their cumulants. These functions are interpreted as the reduced correlation functions
and they are governed by the corresponding hierarchy of the nonlinear evolution equations \cite{B49}-\cite{Ch}.

In this article, we develop an approach to the description of the evolution of a state by means of
reduced distribution functions and reduced correlation functions, which is based on the dynamics of
correlations in a system of hard spheres. It should be emphasized that the generating operators of
solution expansions of the Cauchy problem of the corresponding hierarchies of evolution equations
are induced by the generating operators of an expansion of the solution of the Cauchy problem of
the Liouville hierarchy for correlation functions which are represented by means of the corresponding-order
cumulants of the groups of operators  of the Liouville equations \cite{GG21},\cite{GerRS}.

In the last section of the article, as an application of a developed approach to describing the evolution
of the state of many hard spheres within the framework of the dynamics of correlations, we consider two
issues related to the problem of the rigorous derivation of kinetic equations.
One of them is a method for describing the evolution of the state of many hard spheres within the framework
of the evolution of the state of a typical particle, described by a generalization of the Enskog equation
\cite{GG}, or, in other words, the fundamentals of describing the evolution of correlations by kinetic
equations are discussed.
We remark that the conventional approach to the mentioned problem is based on the construction of the
Boltzmann--Grad asymptotic behavior \cite{G58}-\cite{PG90} of a solution of the BBGKY hierarchy for
reduced distribution functions represented as the series of the perturbation theory for initial data
specified by a one-particle distribution function in the case of absence of correlations (so-called
initial chaos) \cite{CGP97}-\cite{SR12}. In this connection, we also consider a sketch of constructing
the scaling asymptotics of the reduced correlation functions in the low-density limit.


\textcolor{blue!50!black}{\section{Dynamics of correlations of a hard-sphere system}}

\subsection{Preliminaries: dynamics of finitely many hard spheres}
At the initial instant in the space $\mathbb{R}^{3}$ the state of a hard-sphere system of
a non-fixed, i.e. arbitrary but finite average number of identical particles, is described
by the sequence $D(0)=(1,D_{1}(0),\ldots,D_{n}(0),\ldots)$ of the probability distribution
functions $D_{n}(0),\,n\geq1$ defined on the phase space
$\mathbb{R}^{3n}\times(\mathbb{R}^{3n}\setminus \mathbb{W}_n)$ of $n$ hard spheres. Each hard
sphere of a diameter $\sigma>0$ is characterized by the phase space coordinates
$(q_{i},p_{i})\equiv x_{i}\in\mathbb{R}^{3}\times\mathbb{R}^{3},\,i\geq1$ and for configurations
in the space the following inequalities are satisfied: $|q_i-q_j|\geq\sigma,$ $i\neq j\geq1$,
i.e. the set $\mathbb{W}_n\equiv\big\{(q_1,\ldots,q_n)\in\mathbb{R}^{3n}\big||q_i-q_j|<\sigma$
for at least one pair $(i,j):\,i\neq j\in(1,\ldots,n)\big\}$ is the forbidden configuration set.
The nonnegative functions $D_{n}(0)=D_{n}(0,x_1,\ldots,x_n),\,n\geq1,$ that are symmetric with
respect to permutations of the arguments $x_1,\ldots,x_n$, equal to zero on the set of forbidden
configurations $\mathbb{W}_n$ will be assumed to belong to the space of integrable functions
$L^{1}_{n}\equiv L^{1}(\mathbb{R}^{3n}\times\mathbb{R}^{3n})$ equipped with the norm:
$\|f_n\|_{L^{1}_{n}}=\int_{\mathbb{R}^{3n}\times\mathbb{R}^{3n}}dx_1\ldots dx_n|f_n(x_1,\ldots,x_n)|$.

As is known \cite{CGP97}-\cite{Bog}, the evolution of all possible states of a hard-sphere system
is described by the sequence $D(t)=(1,D_{1}(t),\ldots,D_{n}(t),\ldots)$ of the following probability
distribution functions:
\begin{eqnarray}\label{rozv_L}
    &&\hskip-5mm D_{n}(t,x_1,\ldots,x_n)=\begin{cases}
         D_n(0,X_{1}(-t,x_{1},\ldots,x_{n}),\ldots,X_{n}(-t,x_{1},\ldots,x_{n})),\\
         \hskip+45mm\mathrm{if}\,(x_{1},\ldots,x_{n})\in(\mathbb{R}^{3n}\times(\mathbb{R}^{3n}\setminus\mathbb{W}_{n})),\\
         0, \hskip+41mm \mathrm{if}\,(q_{1},\ldots,q_{n})\in\mathbb{W}_{n},
    \end{cases}
\end{eqnarray}
where for $t\in\mathbb{R}$ the function $X_{i}(-t)$ is a phase space trajectory of the $ith$ hard sphere
constructed in \cite{CGP97}, which is defined almost everywhere on the phase space
$\mathbb{R}^{3n}\times(\mathbb{R}^{3n}\setminus \mathbb{W}_n)$, namely, beyond the set $\mathbb{M}_{n}^0$
of zero Lebesgue measure consisting of the phase space points specified by such initial data that
multiple collisions can occur during the evolution, i.e. collisions of more than two particles, more than
one two-particle collision at the same instant and infinite number of collisions within a finite time
interval \cite{CGP97},\cite{PG90}.

On the space $L^{1}_{n}\equiv L^{1}(\mathbb{R}^{3n}\times\mathbb{R}^{3n})$ one-parameter mapping (\ref{rozv_L})
generates an isometric strong continuous group of operators
\begin{eqnarray}\label{Sspher}
   &&S_n(-t,1,\ldots,n)D_n(0,x_1,\ldots,x_n)\doteq D_{n}(t,x_1,\ldots,x_n).
\end{eqnarray}
On the subspace of continuously differentiable functions with compact supports $L_{n,0}^1\subset L^1_n$
for the group of operators (\ref{Sspher}) the Duhamel equation holds
\begin{eqnarray*}
    &&\hskip-5mm S_n(-t,1,\ldots,n)=\prod\limits_{i=1}^{n}S_1(-t,i)+\int\limits_0^t d\tau
       \prod\limits_{i=1}^{n}S_{1}(\tau-t,i)
       \sum\limits_{j_{1}<j_{2}=1}^{n}\mathcal{L}_{\mathrm{int}}^\ast(j_{1},j_{2})S_{n}(-\tau,1,\ldots,n))=\\
    &&\prod\limits_{i=1}^{n}S_1(-t,i)+\int\limits_0^td\tau S_{n}(\tau-t,1,\ldots,n)
       \sum\limits_{j_{1}<j_{2}=1}^{n}\mathcal{L}_{\mathrm{int}}^\ast(j_{1},j_{2})
       \prod\limits_{i=1}^{n}S_{1}(-\tau,i),\nonumber
\end{eqnarray*}
where for $t>0$ the operator $\mathcal{L}_{\mathrm{int}}^\ast(j_{1},j_{2})$ is defined by the formula
\begin{eqnarray}\label{Lint}
     &&\hskip-5mm \mathcal{L}_{\mathrm{int}}^\ast(j_{1},j_{2})f_{n}
        \doteq\sigma^2\int\limits_{\mathbb{S}_{+}^2}d\eta\langle\eta,(p_{j_{1}}-p_{j_{2}})\rangle
        f_n(x_1,\ldots,p_{j_{1}}^*,q_{j_{1}},\ldots,\\
     &&p_{j_{2}}^*,q_{j_{2}},\ldots,x_n)\delta(q_{j_{1}}-q_{j_{2}}+\sigma\eta)-
        f_n(x_1,\ldots,x_n)\delta(q_{j_{1}}-q_{j_{2}}-\sigma\eta)\big).\nonumber
\end{eqnarray}
In definition (\ref{Lint}) the symbol $\langle \cdot,\cdot \rangle$ means a scalar product, the symbol $\delta$ denotes
the Dirac measure, $\mathbb{S}_{+}^{2}\doteq\{\eta\in\mathbb{R}^{3}\big|\left|\eta\right|=1\langle\eta,(p_1-p_{2})\rangle>0\}$
and the pre-collision momenta $p_{i}^\ast,p_{j}^\ast$ are determined by the equalities:
\begin{eqnarray*}\label{momenta}
     &&p_i^\ast\doteq p_i-\eta\left\langle\eta,\left(p_i-p_{j}\right)\right\rangle, \\
     &&p_{j}^\ast\doteq p_{j}+\eta\left\langle\eta,\left(p_i-p_{j}\right)\right\rangle \nonumber.
\end{eqnarray*}
In the case $t<0$, the operator $\mathcal{L}_{\mathrm{int}}(j_{1},j_{2})$ is defined by the corresponding
expression \cite{CGP97}.

Hence the infinitesimal generator $\mathcal{L}_{n}^\ast$ of the group of operators $S_n(t)$
has the following structure
\begin{eqnarray}\label{Lstar}
    &&\mathcal{L}_{n}^\ast(1,\ldots,n) f_{n}\doteq\sum\limits_{j=1}^{n}\mathcal{L}^\ast(j)f_{n}+
        \sum\limits_{j_{1}<j_{2}=1}^{n}\mathcal{L}_{\mathrm{int}}^\ast(j_{1},j_{2})f_{n},
\end{eqnarray}
where the Liouville operator of free motion
$\mathcal{L}^{\ast}(j)\doteq-\langle p_j,\frac{\partial}{\partial q_j}\rangle$ defined on
the subspace $L_{n,0}^1\subset L^1_n$, is denoted by the symbol $\mathcal{L}^{\ast}(j)$.

If $D_n(0)\in L^{1}_{n},\, n\geq 1$, the sequence of distribution functions defined by formula
(\ref{rozv_L}) is a unique solution of the Cauchy problem of the weak
formulation of a sequence of evolution equations for the state known as the Liouville equation:
\begin{eqnarray}
  \label{Liouville}
     &&\frac{\partial}{\partial t}D_n(t)=\mathcal{L}_n^\ast(1,\ldots,n) D_n(t),\\ \nonumber\\
  \label{Liouville_i}
     &&D_n(t)|_{t=0}=D_n(0), \quad n\geq 1.
\end{eqnarray}

Thus, the traditional approach to the description of the evolution of all possible states of
finitely many hard spheres is specified by the Cauchy problem for the sequence of Liouville
equations.

\subsection{Correlation functions}
An alternative approach to the description of states of a hard-sphere system of finitely many
particles is given by means of functions determined by the cluster expansions of the probability
distribution functions. They are  called here as correlation functions (the cumulants of probability
distribution functions).

We introduce the sequence of correlation functions $g(t)=(1,g_{1}(t,x_1),\ldots,g_{s}(t,x_1,\ldots,x_s),\ldots)$
by means of cluster expansions of the probability distribution functions
$D(t)=(1,D_{1}(t,x_1),\ldots,$ $D_{n}(t,x_1,\ldots,x_n),\ldots)$, defined on the set of allowed
configurations $\mathbb{R}^{3n}\setminus \mathbb{W}_n$ as follows:
\begin{eqnarray}\label{D_(g)N}
    &&\hskip-7mm D_{n}(t,x_1,\ldots,x_n)= g_{n}(t,x_1,\ldots,x_n)+\sum\limits_{\mbox{\scriptsize $\begin{array}{c}\mathrm{P}:
    (x_1,\ldots,x_n)=\bigcup_{i}X_{i},\\|\mathrm{P}|>1 \end{array}$}}
        \prod_{X_i\subset \mathrm{P}}g_{|X_i|}(t,X_i),\,\,\,\,n\geq1,
\end{eqnarray}
where ${\sum\limits}_{\mathrm{P}:(x_1,\ldots,x_n)=\bigcup_{i} X_{i},\,|\mathrm{P}|>1}$ is the sum
over all possible partitions $\mathrm{P}$ of the set of the arguments $(x_1,\ldots,x_n)$ into
$|\mathrm{P}|>1$ nonempty mutually disjoint subsets $X_i\subset(x_1,\ldots,x_n)$.

On the set $\mathbb{R}^{3n}\setminus \mathbb{W}_n$ solutions of recursion relations (\ref{D_(g)N})
are given by the following expansions:
\begin{eqnarray}\label{gfromDFB}
   &&\hskip-7mm g_{s}(t,x_1,\ldots,x_s)=D_{s}(t,x_1,\ldots,x_s)+\\
   &&\sum\limits_{\mbox{\scriptsize $\begin{array}{c}\mathrm{P}:(x_1,\ldots,x_s)=
       \bigcup_{i}X_{i},\\|\mathrm{P}|>1\end{array}$}}(-1)^{|\mathrm{P}|-1}(|\mathrm{P}|-1)!\,
       \prod_{X_i\subset \mathrm{P}}D_{|X_i|}(t,X_i), \quad s\geq1.\nonumber
\end{eqnarray}
The structure of expansions (\ref{gfromDFB}) is such that the correlation functions can be treated
as cumulants (semi-invariants) of the probability distribution functions (\ref{rozv_L}).

Thus, correlation functions (\ref{gfromDFB}) are to enable to describe of the evolution of states
of finitely many hard spheres by the equivalent method in comparison with probability distribution
functions (\ref{rozv_L}), namely within the framework of dynamics of correlations.

If the initial state is specified by the sequence $g(0)=(1,g_{1}^{0}(x_1),\ldots,g_{n}^{0}(x_1,\ldots,$ $x_n),\ldots)$,
of correlation functions $g_{n}^{0}\in L^{1}_{n},\,n\geq1,$ then the evolution of all possible states,
i.e. the sequence $g(t)=(1,$ $g_{1}(t,x_1),\ldots,g_{s}(t,x_1,\ldots,x_s),\ldots)$ of the correlation
functions $g_{s}(t),\,s\geq1$, is determined by the following expansions:
\begin{eqnarray}\label{ghs}
   &&\hskip-8mm g_{s}(t,x_1,\ldots,x_s)=\sum\limits_{\mathrm{P}:\,(x_1,\ldots,x_s)=\bigcup_j X_j}
      \mathfrak{A}_{|\mathrm{P}|}(t,\{\widehat{X}_1\},\ldots,\{\widehat{X}_{|\mathrm{P}|}\})
      \prod_{X_j\subset \mathrm{P}}g_{|X_j|}^{0}(X_j),\quad s\geq1,
\end{eqnarray}
where the symbol $\sum_{\mathrm{P}:\,(x_1,\ldots,x_s)=\bigcup_j X_j}$ is the sum over all possible
partitions $\mathrm{P}$ of the set $(x_1,\ldots,x_s)$ into $|\mathrm{P}|$ nonempty mutually disjoint
subsets $X_j$, the symbol $\widehat{X}$ means the set of indexes of the set $X$ of phase space
coordinates and the set $(\{\widehat{X}_1\},\ldots,\{\widehat{X}_{|\mathrm{P}|}\})$ consists of
elements that are subsets $\widehat{X}_j\subset (1,\ldots,s)$, i.e.
$|(\{\widehat{X}_1\},\ldots,\{\widehat{X}_{|\mathrm{P}|}\})|=|\mathrm{P}|$.
The generating operator $\mathfrak{A}_{|\mathrm{P}|}(t)$ in expansions (\ref{ghs}) is the
$|\mathrm{P}|th$-order cumulant of the groups of operators (\ref{Sspher}) which is defined
by the expansion
\begin{eqnarray}\label{cumulantP}
   &&\hskip-7mm \mathfrak{A}_{|\mathrm{P}|}(t,\{\widehat{X}_1\},\ldots,\{\widehat{X}_{|\mathrm{P}|}\})\doteq\\
   && \sum\limits_{\mathrm{P}^{'}:\,(\{\widehat{X}_1\},\ldots,\{\widehat{X}_{|\mathrm{P}|}\})=
      \bigcup_k Z_k}(-1)^{|\mathrm{P}^{'}|-1}({|\mathrm{P}^{'}|-1})!
      \prod\limits_{Z_k\subset\mathrm{P}^{'}}S_{|\theta(Z_{k})|}(-t,\theta(Z_{k})),\nonumber
\end{eqnarray}
where $\theta$ is the declusterization mapping:
$\theta(\{\widehat{X}_1\},\ldots,\{\widehat{X}_{|\mathrm{P}|}\})\doteq(1,\ldots,s)$.
The simplest examples of correlation functions (\ref{ghs}) are given as follows:
\begin{eqnarray*}
   &&g_{1}(t,x_1)=\mathfrak{A}_{1}(t,1)g_{1}^{0}(x_1),\\
   &&g_{2}(t,x_1,x_2)=\mathfrak{A}_{1}(t,\{1,2\})g_{2}^{0}(x_1,x_2)+
     \mathfrak{A}_{2}(t,1,2)g_{1}^{0}(x_1)g_{1}^{0}(x_2).
\end{eqnarray*}

The structure of expansions (\ref{ghs}) is established as a result of the permutation
of the terms of cumulant expansions (\ref{gfromDFB}) for correlation functions and cluster
expansions (\ref{D_(g)N}) for initial probability distribution functions (\ref{Sspher}).
Thus, the cumulant origin of correlation functions induces the cumulant structure of their
dynamics (\ref{ghs}).

In particular, in the absence of correlations between hard spheres at the initial moment (initial state
satisfying the chaos condition \cite{CGP97},\cite{Sp91}) the sequence of the initial correlation functions
on allowed configurations has the form $g^{(c)}(0)=(0,g_{1}^{0}(x_1),0,\ldots,0,\ldots)$.
In terms of a sequence of the probability distribution functions the chaos condition means that we start from 
an initial datum of the form
$D^{(c)}(0)=(1,D_{1}^0(x_1),D_{1}^0(x_1)D_{1}^0(x_2)\mathcal{X}_{\mathbb{R}^{6}\setminus \mathbb{W}_2},\ldots,
$ $\prod^n_{i=1}D_{1}^0(x_i)\mathcal{X}_{\mathbb{R}^{3n}\setminus \mathbb{W}_n},\ldots)$,
where the function $\mathcal{X}_{\mathbb{R}^{3n}\setminus \mathbb{W}_{n}}$ is the Heaviside step function of
allowed configurations of $n$ hard spheres.
In this case for $(x_1,\ldots,x_s)\in\mathbb{R}^{3s}\times(\mathbb{R}^{3s}\setminus \mathbb{W}_s)$
expansions (\ref{ghs}) are represented as follows:
\begin{eqnarray}\label{gth}
   &&g_{s}(t,x_1,\ldots,x_s)=\mathfrak{A}_{s}(t,1,\ldots,s)\,\prod\limits_{i=1}^{s}g_{1}^{0}(x_i)
         \mathcal{X}_{\mathbb{R}^{3s}\setminus \mathbb{W}_s},\quad s\geq1,
\end{eqnarray}
where the generating operator $\mathfrak{A}_{s}(t)$ of this expansion is the $sth$-order cumulant
of groups of operators (\ref{Sspher}) defined by the expansion
\begin{eqnarray}\label{cumcp}
   &&\hskip-8mm\mathfrak{A}_{s}(t,1,\ldots,s)=\sum\limits_{\mathrm{P}:\,(1,\ldots,s)=
       \bigcup_i X_i}(-1)^{|\mathrm{P}|-1}({|\mathrm{P}|-1})!
      \prod\limits_{X_i\subset\mathrm{P}}S_{|X_i|}(-t,X_i),
\end{eqnarray}
with notations accepted in formula (\ref{ghs}). From the structure of series (\ref{gth})
it is clear that in case of the absence of correlations at the initial instant the correlations generated
by the dynamics of a system of hard spheres are completely determined by cumulants (\ref{cumcp}) of the
groups of operators of the Liouville equation (\ref{Liouville}).


We note that in the case of initial data $g^{(c)}(0)$ expansions (\ref{gth}) can be rewritten in another
representation that explains their physical meaning. Indeed, for $n=1$ we have
\begin{eqnarray*}
     &&g_{1}(t,x_1)=\mathfrak{A}_{1}(t,1)g_{1}^0(x_1)=g_{1}^0(p_1,q_1-p_1t).
 \end{eqnarray*}
Then, according to formula (\ref{gth}) and the definition of the first-order cumulant $\mathfrak{A}_{1}(t)=S_1(-t)$,
and its inverse group of operators $S_1^{-1}(-t)=S_1(t)$, we express the correlation functions
$g_{s}(t),$ $s\geq 2$, in terms of the one-particle correlation function $g_{1}(t)$.
Therefore, for $s\geq2$ expansions (\ref{gth}) are represented in the following form:
\begin{eqnarray*}
     &&g_{s}(t,x_1,\ldots,x_s)=\widehat{\mathfrak{A}}_{s}(t,1,\ldots,s)\,\prod_{i=1}^{s}\,g_{1}(t,x_i),\quad s\geq 2,
\end{eqnarray*}
where $\widehat{\mathfrak{A}}_{s}(t,1,\ldots,s)$ is the $s$-order cumulant (\ref{cumcp})
of the scattering operators
\begin{eqnarray*}
     &&\widehat{S}_n(t,1,\ldots,n)\doteq
        S_{n}(-t,1,\ldots,n)\mathcal{X}_{\mathbb{R}^{3n}\setminus \mathbb{W}_n}\prod_{i=1}^{n}S_{1}(t,i),\quad n\geq1.
\end{eqnarray*}
On subspace $L^{1}_{0,n}\subset L^{1}_{n}$ the generator of the scattering operator $\widehat{S}_n(t,1,\ldots,n)$
is determined by the operator:
$\frac{d}{dt}\widehat{S}_n(t,1,\ldots,n)\mid_{t=0}=\sum_{j_{1}<j_{2}=1}^{n}\mathcal{L}_{\mathrm{int}}^\ast(j_{1},j_{2}),$
where for $t\geq0$ the operator $\mathcal{L}_{\mathrm{int}}^\ast(j_{1},j_{2})$ is defined according to formula (\ref{Lint}).

If $g^0_{n}\in L^{1}_{n},\,n\geq1$, one-parameter mapping (\ref{ghs}) generates strong continuous
group of nonlinear operators
\begin{eqnarray}\label{Sspherc}
   &&\mathcal{G}(t;1,\ldots,s\mid g(0))\doteq g_{s}(t,x_1,\ldots,x_s),
\end{eqnarray}
and it is bounded, and the following estimate holds:
$\big\|\mathcal{G}(t;1,\ldots,s\mid g)\big\|_{L^{1}_{s}}\leq s!c^{s},$
where $c\equiv \max(1,$ $\max_{\mathrm{P}:\,(1,\ldots,s)=\bigcup_i X_i}\|g_{|X_{i}|}\|_{L^{1}_{|X_{i}|}})$.
For $g_{n}\in L^{1}_{n,0},\,n\geq1$, the infinitesimal generator of this group of nonlinear operators has
the following structure
\begin{eqnarray}\label{LL}
   &&\hskip-5mm \mathcal{L}(1,\ldots,s\mid g)\doteq\mathcal{L}^{\ast}_{s}(1,\ldots,s)g_{s}(x_1,\ldots,x_s)+\\
   &&\sum\limits_{\mathrm{P}:\,(x_1,\ldots,x_s)=X_{1}\bigcup X_2}\,\sum\limits_{i_{1}\in \widehat{X}_{1}}
      \sum\limits_{i_{2}\in \widehat{X}_{2}}\mathcal{L}_{\mathrm{int}}^{\ast}(i_{1},i_{2})
      g_{|X_{1}|}(X_{1})g_{|X_{2}|}(X_{2}),\nonumber
\end{eqnarray}
where we used the notation adopted above in expansions (\ref{ghs}).


\subsection{The Liouville hierarchy}
If $g_{s}^{0}\in L^{1}_{s},\,s\geq1,$ then for $t\in\mathbb{R}$ the sequence of correlation
functions (\ref{ghs}) defined on the set of allowed configurations is a unique solution of
the Cauchy problem of the weak formulation of the Liouville hierarchy:
\begin{eqnarray}\label{Lh}
   &&\hskip-8mm\frac{\partial}{\partial t}g_{s}(t,x_1,\ldots,x_s)=
      \mathcal{L}^{\ast}_{s}(1,\ldots,s)g_{s}(t,x_1,\ldots,x_s)+\\
   &&\sum\limits_{\mathrm{P}:\,(x_1,\ldots,x_s)=X_{1}\bigcup X_2}\,\sum\limits_{i_{1}\in \widehat{X}_{1}}
      \sum\limits_{i_{2}\in \widehat{X}_{2}}\mathcal{L}_{\mathrm{int}}^{\ast}(i_{1},i_{2})
      g_{|X_{1}|}(t,X_{1})g_{|X_{2}|}(t,X_{2}), \nonumber\\
   \nonumber\\
 \label{Lhi}
   &&\hskip-8mm g_{s}(t,x_1,\ldots,x_s)\big|_{t=0}=g_{s}^{0}(x_1,\ldots,x_s),\quad s\geq1,
\end{eqnarray}
where ${\sum\limits}_{\mathrm{P}:\,(x_1,\ldots,x_s)=X_{1}\bigcup X_2}$ is the sum over all possible
partitions $\mathrm{P}$ of the set $(x_1,\ldots,x_s)$ into two nonempty mutually disjoint subsets
$X_1$ and $X_2$, the symbol $\widehat{X}_i$ means the set of indexes of the set $X_i$ of phase space
coordinates and the operator $\mathcal{L}^{\ast}_{s}$ is defined on the subspace
$L^{1}_{0}\subset L^{1}$ by formula (\ref{Lstar}). It should be noted that the Liouville hierarchy
(\ref{Lh}) is the evolution recurrence equations set.

For $t\geq0$ we give a few examples of recurrence equations set (\ref{Lh}) for a system of hard spheres:
\begin{eqnarray*}
   &&\hskip-7mm\frac{\partial}{\partial t}g_{1}(t,x_1)=
     -\langle p_1,\frac{\partial}{\partial q_1}\rangle g_{1}(t,x_1),\\
   &&\hskip-7mm\frac{\partial}{\partial t}g_{2}(t,x_1,x_2)=
     -\sum\limits_{j=1}^{2}\langle p_j,\frac{\partial}{\partial q_j}\rangle g_{2}(t,x_1,x_2)+ \\
   && \sigma^2\int\limits_{\mathbb{S}_{+}^2}d\eta\langle\eta,(p_{1}-p_{2})\rangle
        \big(g_2(t,q_1,p_{1}^\ast,q_{2},p_{2}^\ast)\delta(q_{1}-q_{2}+\sigma\eta)-\\
   &&g_2(t,x_1,x_2)\delta(q_{1}-q_{2}-\sigma\eta)\big)+\\
   &&\sigma^2\int\limits_{\mathbb{S}_{+}^2}d\eta\langle\eta,(p_{1}-p_{2})\rangle
        \big(g_1(t,q_1,p_{1}^\ast)g_1(t,q_{2},p_{2}^\ast)\delta(q_{1}-q_{2}+\sigma\eta)-\\
   &&g_1(t,x_1)g_1(t,x_2)\delta(q_{1}-q_{2}-\sigma\eta)\big),
\end{eqnarray*}
where it was used notations accepted above in definition (\ref{Lint}).

We note that because the Liouville hierarchy (\ref{Lh}) is the recurrence evolution equations
set, we can construct a solution of the Cauchy problem (\ref{Lh}),(\ref{Lhi}), integrating
each equation of the hierarchy as the inhomogeneous Liouville equation. For example, as a result
of the integration of the first two equations of the Liouville hierarchy (\ref{Lh}), we obtain
the following equalities:
\begin{eqnarray*}
    &&\hskip-5mm g_{1}(t,x_1)=S_{1}(-t,1)g_{1}^{0}(x_1),\\
    &&\hskip-5mm g_{2}(t,1,2)=S_{2}(-t,1,2)g_{2}^0(x_1,x_2)+\\
    &&   \int\limits_{0}^{t}dt_{1}S_{2}(t_{1}-t,1,2)\mathcal{L}^\ast_{\mathrm{int}}(1,2)
       S_{1}(-t_{1},1)S_{1}(-t_{1},2)g_{1}^0(x_1)g_{1}^0(x_2).
\end{eqnarray*}
Then for the corresponding term on the right-hand side of the second equality, an analog of the
Duhamel equation holds
\begin{eqnarray*}
    &&\int\limits_{0}^{t}dt_{1}S_{2}(t_{1}-t,1,2)\mathcal{L}^\ast_{\mathrm{int}}(1,2)
       S_{1}(-t_{1},1)S_{1}(-t_{1},2)=\\
    &&-\int\limits_{0}^{t}dt_{1}\frac{d}{dt_{1}}\big(S_{2}(t_{1}-t,1,2)
       S_{1}(-t_{1},1)S_{1}(-t_{1},2)\big)=\nonumber\\
    &&S_{2}(-t,1,2)-S_{1}(-t,1)S_{1}(-t,2)=\mathfrak{A}_{2}(t,1,2),\nonumber
\end{eqnarray*}
where $\mathfrak{A}_{2}(t)$ is the second-order cumulant (\ref{cumcp}) of groups of operators
(\ref{Sspher}). As a result of similar transformations for $s>2$, the solution of the Cauchy
problem (\ref{Lh}),(\ref{Lhi}), constructed by an iterative procedure, is represented in the
form of expansions (\ref{ghs}).


The following statement is true. 

For $t\in\mathbb{R}$ a unique solution of the Cauchy problem of the Liouville hierarchy
(\ref{Lh}),(\ref{Lhi}) is represented by a sequence of expansions (\ref{ghs}).
For $g_{n}^0\in L^{1}_{n,0}\subset L^{1}_{n},\,n\geq1$, a sequence of functions (\ref{ghs})
is a classical solution and for arbitrary initial data $g_{n}^0\in L^{1}_{n},\,n\geq1$, one has
a generalized solution.

The proof of the theorem is similar to the proof of the existence theorem for the BBGKY hierarchy
in the space of sequences of integrable functions \cite{CGP97},\cite{GerRS}. Indeed, if the initial
data is $g_{n}^0\in L^{1}_{n,0},\,n\geq1$, then the infinitesimal generator of the group of nonlinear
operators (\ref{Sspherc}) coincides with the operator (\ref{LL}) and hence the Cauchy problem
(\ref{Lh}),(\ref{Lhi}) has a classical (strong) solution (\ref{ghs}).


We remark that a steady solution of the Liouville hierarchy (\ref{Lh}) is a sequence of
the Ursell functions on the allowed configurations of hard spheres, i.e. it is the sequence
$g^{(eq)}=(0,e^{-\beta \frac{p^2_1}{2}},0,\ldots,)$, where $\beta$ is a parameter
inversely proportional to temperature.

Finally, we emphasize that the dynamics of correlations, that is, the fundamental equations
(\ref{Lh}) describing the evolution of correlations of states of hard spheres, can be used
as a basis for describing the evolution of the state of both a finite and an infinite number
of hard spheres instead of the Liouville equations for the state (\ref{Liouville}).
Further, we establish that the constructed dynamics of correlation underlies the description
of the dynamics of infinitely many hard-spheres governed by the BBGKY hierarchies for reduced
distribution functions or reduced correlation functions.

\textcolor{blue!50!black}{\section{Propagation of correlations in a hard-sphere system}}

\subsection{Evolution of states described by the dynamics of correlations:\\ reduced distribution functions}
As is known, an equivalent approach adapted to describing the evolution of states
of systems of both finite and infinite number of hard spheres is to describe a state
by means of a sequence of so-called reduced distribution functions (marginals) governed
by the BBGKY hierarchy \cite{CGP97}. In what follows, we outline an approach to describing
the evolution of a state using both reduced distribution functions and reduced correlation
functions, based on the dynamics of correlations in a system of hard spheres governed by
the Liouville hierarchy for correlation functions (\ref{Lh}).

For a hard-sphere system of a non-fixed, i.e. arbitrary but finite average number of identical
particles, the reduced distribution functions are defined by means of probability distribution
functions as follows \cite{CGP97}:
\begin{eqnarray}\label{ms}
     &&\hskip-12mm F_{s}(t,x_1,\ldots,x_s)\doteq(I,D(t))^{-1}\sum\limits_{n=0}^{\infty}\frac{1}{n!}
         \,\int\limits_{(\mathbb{R}^{3}\times\mathbb{R}^{3})^{n}}dx_{s+1}\ldots dx_{s+n}
         \,D_{s+n}(t,x_1,\ldots,x_{s+n}),\\
    && \hskip-12mm   s\geq1,\nonumber
\end{eqnarray}
where the normalizing factor $(I,D(t))\doteq\sum_{n=0}^{\infty}\frac{1}{n!}
\int_{(\mathbb{R}^{3}\times\mathbb{R}^{3})^{n}}dx_{1}\ldots dx_{n}D_{n}(t,x_1,\ldots,x_n)$
is a grand canonical partition function.
The possibility of redefining of the reduced distribution functions naturally arises as a result
of dividing the series in expression (\ref{ms}) by the series of the normalization factor.

A definition of reduced distribution functions equivalent to definition (\ref{ms})
is formulated on the basis of correlation functions (\ref{ghs}) of a system of hard
spheres by means of the following series expansion:
\begin{eqnarray}\label{FClusters}
    &&\hskip-12mm  F_{s}(t,x_1,\ldots,x_s)\doteq\sum\limits_{n=0}^{\infty}\frac{1}{n!}\,
       \int\limits_{(\mathbb{R}^{3}\times\mathbb{R}^{3})^{n}}dx_{s+1}\ldots dx_{s+n}\,
       g_{1+n}(t,\{x_1,\ldots,x_s\},x_{s+1},\ldots,x_{s+n}), \\
    && \hskip-12mm   s\geq1,\nonumber
\end{eqnarray}
where on the set of allowed configurations $\mathbb{R}^{3(s+n)}\setminus \mathbb{W}_{s+n}$ the correlation
functions of clusters of hard spheres $g_{1+n}(t), n\geq0,$ are determined by the expansions:
\begin{eqnarray}\label{rozvL-Nclusters}
    &&\hskip-12mm g_{1+n}(t,\{x_1,\ldots,x_s\},x_{s+1},\ldots,x_{s+n})=\\
    &&\sum_{\mbox{\scriptsize$\begin{array}{c}\mathrm{P}:(\{x_1,\ldots,x_s\},\\
       x_{s+1},\ldots,x_{s+n})=\bigcup_i X_i\end{array}$}}
       \mathfrak{A}_{|\mathrm{P}|}\big(t,\{\theta(\widehat{X}_1)\},\ldots,\{\theta(\widehat{X}_{|\mathrm{P}|})\}\big)
       \prod_{X_i\subset \mathrm{P}}g_{|X_i|}^0(X_i), \quad n\geq0.\nonumber
\end{eqnarray}
We remind that in expansions (\ref{rozvL-Nclusters}) the symbol
$\sum_{\mathrm{P}:(\{x_1,\ldots,x_s\},x_{s+1},\ldots,x_{s+n})=\bigcup_i X_i}$
means the sum over all possible partitions $\mathrm{P}$ of the set $(\{x_1,\ldots,x_s\},x_{s+1},\ldots,$ $x_{s+n})$
into nonempty mutually disjoint subsets $X_i$, and the generating operator $\mathfrak{A}_{|\mathrm{P}|}(t)$
is the $|\mathrm{P}|th$-order cumulant (\ref{cumulantP}) of the groups of operators (\ref{Sspher}).

On allowed configurations the correlation functions of particle clusters in series (\ref{FClusters}),
i.e. the functions $g_{1+n}(t,\{x_{1},\ldots,x_{s}\},$ $x_{s+1},\ldots,x_{s+n}),\,n\geq0$, are defined as solutions
of generalized cluster expansions of a sequence of solutions of the Liouville equations (\ref{Liouville}):
\begin{eqnarray}\label{a8}
    &&\hskip-7mm D_{s+n}(t,x_{1},\ldots,x_{s+n})=\sum_{\mbox{\scriptsize$\begin{array}{c}\mathrm{P}:(\{x_1,\ldots,x_s\},\\
      x_{s+1},\ldots,x_{s+n})=\bigcup_i X_i\end{array}$}}
      \prod_{X_i\subset\mathrm{P}}g_{|X_i|}(t,X_i),\quad s\geq1,\,n\geq0,
\end{eqnarray}
namely,
\begin{eqnarray*}
    &&\hskip-7mm g_{1+n}(t,\{x_{1},\ldots,x_{s}\},x_{s+1},\ldots,x_{s+n})=\\
    &&\sum_{\mbox{\scriptsize$\begin{array}{c}\mathrm{P}:(\{x_1,\ldots,x_s\},\\
       x_{s+1},\ldots,x_{s+n})=\bigcup_i X_i\end{array}$}}
      (-1)^{|\mathrm{P}|-1}(|\mathrm{P}| -1)!\,\prod_{X_i\subset\mathrm{P}}D_{|\theta(X_{i})|}(t,\theta(X_{i})),
      \quad s\geq1,\,n\geq0,\nonumber
\end{eqnarray*}
where $\theta$ is the declusterization mapping defined in formula (\ref{cumulantP}), the probability
distribution function $D_{|\theta(X_{i})|}(t,\theta(X_{i}))$ is solution (\ref{rozv_L}) of the Liouville
equations (\ref{Liouville}).

The correlation functions of particle clusters satisfy the Liouville hierarchy of evolution equations
with the following generator
\begin{eqnarray}\label{LcL}
   &&\hskip-12mm \mathcal{L}(\{1,\ldots,s\},s+1,\ldots,s+n\mid \mathfrak{d}_{\{Y\}}g(t))\doteq\\
    &&\hskip+5mm \mathcal{L}^{\ast}_{s+n}(1,\ldots,s+n)g_{1+n}(t,X)+\nonumber \\
   &&\hskip+5mm \sum\limits_{\mathrm{P}:\,X=X_{1}\bigcup X_2}\,
      \sum\limits_{i_{1}\in\theta(\widehat{X}_{1})}
      \sum\limits_{i_{2}\in\theta(\widehat{X}_{2})}\mathcal{L}_{\mathrm{int}}^{\ast}(i_{1},i_{2})
      g_{|X_{1}|}(t,X_{1})g_{|X_{2}|}(t,X_{2}), \quad n\geq0,\nonumber
\end{eqnarray}
where $X\equiv(\{Y\},x_{s+1},\ldots,x_{s+n})\equiv(\{x_1,\ldots,x_s\},x_{s+1},\ldots,x_{s+n})$, the sequence 
of solutions of generalized cluster expansions (\ref{a8}) is denoted by means of the mapping
\begin{eqnarray*}
   &&\hskip-5mm(\mathfrak{d}_{\{Y\}}g)_{n}(x_1,\ldots,x_n)\doteq
        g_{1+n}(\{x_1,\ldots,x_s\},x_{s+1},\ldots,x_{s+n}),\quad n\geq0,
\end{eqnarray*}
and we also used the notations adopted above in expansion (\ref{ghs}).

We note that on the allowed configurations correlation functions of hard-sphere clusters can be expressed 
through correlation functions of hard spheres (\ref{ghs}) by the following relations:
\begin{eqnarray}\label{gClusters}
  &&\hskip-12mm g_{1+n}(t,\{x_1,\ldots,x_s\},x_{s+1},\ldots,x_{s+n})=
     \sum_{\mbox{\scriptsize$\begin{array}{c}\mathrm{P}:(\{x_1,\ldots,x_s\},\\
      x_{s+1},\ldots,x_{s+n})=\bigcup_i X_i\end{array}$}}(-1)^{|\mathrm{P}|-1}(|\mathrm{P}|-1)!\times\\
  &&\prod_{X_i\subset\mathrm{P}}\,\sum\limits_{\mathrm{P'}:\,\theta(X_{i})=\bigcup_{j_i} Z_{j_i}}\,
      \prod_{Z_{j_i}\subset\mathrm{P'}}g_{|Z_{j_i}|}(t,Z_{j_i}), \quad n\geq0.\nonumber
\end{eqnarray}
In particular case $n=0$, i.e. the correlation function of a cluster of the $s$ hard spheres,
these relations take the form
\begin{eqnarray*}
  &&g_{1+0}(t,\{x_1,\ldots,x_s\})=\sum\limits_{\mathrm{P}:\,\theta(\{x_1,\ldots,x_s\})=\bigcup_{i} X_{i}}\,
      \prod_{X_{i}\subset \mathrm{P}}g_{|X_{i}|}(t,X_{i}).\nonumber
\end{eqnarray*}

As a consequence of these relations, for the initial state satisfying the chaos condition,
from (\ref{rozvL-Nclusters}) the following generalization of expansions (\ref{gth}) holds:
\begin{eqnarray}\label{gcph}
   &&\hskip-12mm g_{s+n}(t,\{x_1,\ldots,x_s\},x_{s+1},\ldots,x_{s+n})=\\
   && \mathfrak{A}_{1+n}(t,\{1,\ldots,s\},s+1,\ldots,s+n)\,
      \prod\limits_{i=1}^{s+n}g_{1}^{0}(x_i)\mathcal{X}_{\mathbb{R}^{3(s+n)}\setminus\mathbb{W}_{s+n}},
      \quad s\geq1,\,n\geq0.\nonumber
\end{eqnarray}

As we noted above, the possibility of the description of the evolution of a state based on the dynamics
of correlations (\ref{FClusters}) occurs naturally in consequence of dividing the series of expressions
(\ref{ms}) by the series of the normalizing factor. To provide evidence of this statement, we will
introduce the necessary concepts and prove the validity of some auxiliary equalities.

On sequences of functions $f,\widetilde{f}\in L^{1}\oplus_{n=0}^\infty L^{1}_n$ we define the following
$\ast$-product  \cite{R69}
\begin{eqnarray}\label{Product}
    (f\ast\widetilde{f})_{s}(x_1,\ldots,x_s)=\sum\limits_{Z\subset (x_1,\ldots,x_s)}\,f_{|Z|}(Z)
        \,\widetilde{f}_{s-|Z|}((x_1,\ldots,x_s)\setminus Z),
\end{eqnarray}
where $\sum_{Z\subset(x_1,\ldots,x_s)}$ is the sum over all subsets $Z$ of the set $(x_1,\ldots,x_s)$.
Using the definition of the $\ast$-product (\ref{Product}), we introduce the mapping
${\mathbb E}\mathrm{xp}_{\ast}$ and the inverse mapping ${\mathbb L}\mathrm{n}_{\ast}$ on sequences
$h=(0,h_1(x_1),\ldots,h_n(x_1,\ldots,x_n),\ldots)$ of functions $h_n\in L^{1}_n$ by the expansions:
\begin{eqnarray}\label{circledExp}
   &&\hskip-7mm({\mathbb E}\mathrm{xp}_{\ast}\,h)_{s}(x_1,\ldots,x_s)=\big(\mathbb{I}+
      \sum\limits_{n=1}^{\infty}\frac{h^{\ast n}}{n!}\big)_{s}(x_1,\ldots,x_s)=\\
   &&\hskip+12mm\delta_{s,0}+\sum\limits_{\mathrm{P}:\,(x_1,\ldots,x_s)=\bigcup_{i}X_{i}}\,
      \prod_{X_i\subset \mathrm{P}}h_{|X_i|}(X_i),\nonumber
\end{eqnarray}
where we used the notations accepted in formula (\ref{D_(g)N}), $\mathbb{I}=(1,0,\ldots,0,\ldots)$
and $\delta_{s,0}$ is the Kronecker symbol, and respectively,
\begin{eqnarray}\label{circledLn}
   &&\hskip-7mm({\mathbb L}\mathrm{n}_{\ast}(\mathbb{I}+h))_{s}(x_1,\ldots,x_s)=
       \big(\sum\limits_{n=1}^{\infty} (-1)^{n-1}\,\frac{h^{\ast n}}{n}\big)_{s}(x_1,\ldots,x_s)=\\
   &&\hskip+12mm \sum\limits_{\mathrm{P}:\,(x_1,\ldots,x_s)=\bigcup_{i}X_{i}}(-1)^{|\mathrm{P}|-1}(|\mathrm{P}|-1)!\,
       \prod_{X_i\subset\mathrm{P}}h_{|X_i|}(X_i).\nonumber
\end{eqnarray}
Therefore in terms of sequences of functions recursion relations (\ref{D_(g)N}) are rewritten
in the form
\begin{eqnarray*}\label{DtoGcircledStar}
    &&\hskip-7mm D(t)={\mathbb E}\mathrm{xp}_{\ast}\,\,g(t),
\end{eqnarray*}
where $D(t)=\mathbb{I}+(0,D_1(t,x_1),\ldots,D_n(t,x_1,\ldots,x_n),\ldots)$. As a result, we get
\begin{eqnarray*}
    &&\hskip-7mm g(t)={\mathbb L}\mathrm{n}_{\ast}\,\,D(t).
\end{eqnarray*}

Thus, according to definition (\ref{Product}) of the $\ast$-product and mapping (\ref{circledLn}),
in the component-wise form solutions of recursion relations (\ref{D_(g)N}) are represented by
expansions (\ref{gfromDFB}).

For arbitrary $f=(f_{0},f_{1},\ldots,f_{n},\ldots)\in L^{1}$ and the set $Y\equiv(x_1,\ldots,x_s)$
we define the linear mapping $\mathfrak{d}_{Y}:f\rightarrow \mathfrak{d}_{Y}f$, by the formula
\begin{eqnarray}\label{oper_d}
   &&(\mathfrak{d}_{Y}f)_{n}(x_1,\ldots,x_n)\doteq f_{s+n}(x_1,\ldots,x_s,x_{s+1},\ldots,x_{s+n}),
       \quad n\geq0.
\end{eqnarray}
For the set $\{Y\}$ consisting of the one element $Y=(x_1,\ldots,x_s)$, we have, respectively
\begin{eqnarray}\label{oper_c}
   &&\hskip-5mm(\mathfrak{d}_{\{Y\}}f)_{n}(x_1,\ldots,x_n)\doteq
        f_{1+n}(\{x_1,\ldots,x_s\},x_{s+1},\ldots,x_{s+n}),\quad n\geq0.
\end{eqnarray}
On sequences $\mathfrak{d}_{Y}f$ and $\mathfrak{d}_{Y'}\widetilde{f}$ we introduce the $\ast$-product
\begin{eqnarray*}
    &&(\mathfrak{d}_{Y}f\ast\mathfrak{d}_{Y'}\widetilde{f})_{|X|}(X)\doteq
       \sum\limits_{Z\subset X}f_{|Z|+|Y|}(Y,Z)\,\widetilde{f}_{|X\backslash Z|+|Y'|}(Y',X\setminus Z),
\end{eqnarray*}
where $X,Y,Y'$ are the sets, which characterize clusters of hard spheres, and $\sum_{Z\subset X}$
is the sum over all subsets $Z$ of the set $X$. In particular case $Y=\emptyset,\,Y'=\emptyset$,
this definition reduces to definition of $\ast$-product (\ref{Product}).

Let us establish some properties of introduced mappings (\ref{circledExp}) and (\ref{oper_c}).

If $f_{n}\in L^{1}_n,\,n\geq 1$ for the sequences $f=(0,f_{1},\ldots,f_{n},\ldots)$, according to
definitions of mappings (\ref{circledExp}) and (\ref{oper_c}), the following equality holds
\begin{eqnarray}\label{d_gamma}
    &&\mathfrak{d}_{\{Y\}}\mathbb{E}\mathrm{xp}_{\ast}f=
      \mathbb{E}\mathrm{xp}_{\ast}f\ast\mathfrak{d}_{\{Y\}}f,
\end{eqnarray}
and for mapping (\ref{oper_d}) respectively
\begin{eqnarray*}
    &&\mathfrak{d}_{Y}\mathbb{E}\mathrm{xp}_{\ast}f=
       \mathbb{E}\mathrm{xp}_{\ast} f\ast\sum\limits_{\mathrm{P}:\,Y=\bigcup_i X_{i}}
       \mathfrak{d}_{X_1}f\ast\ldots\ast \mathfrak{d}_{X_{|\mathrm{P}|}}f,
\end{eqnarray*}
where ${\sum\limits}_{\mathrm{P}:\,Y=\bigcup_i X_{i}}$ is the sum over all possible partitions
$\mathrm{P}$ of the set $Y\equiv(x_1,\ldots,x_s)$ into $|\mathrm{P}|$ nonempty mutually disjoint
subsets $X_i\subset Y$.

Hence in terms of mappings (\ref{oper_d}) and (\ref{oper_c}) generalized cluster expansions (\ref{a8})
take the form
\begin{eqnarray}\label{gcea}
    &&\mathfrak{d}_{Y}D(t)=\mathfrak{d}_{\{Y\}}{\mathbb E}\mathrm{xp}_{\ast}\,\,g(t).
\end{eqnarray}

On sequences of functions $f\in L^{1}=\oplus_{n=0}^\infty L^{1}_n$ we also define the analogue
of the annihilation operator
\begin{eqnarray}\label{a}
    &&(\mathfrak{a}f)_{n}(x_1,\ldots,x_n)=
      \int\limits_{\mathbb{R}^3\times\mathbb{R}^3}dx_{n+1}f_{n+1}(x_1,\ldots,x_n,x_{n+1}).
\end{eqnarray}
Then for sequences $f,\widetilde{f}\in L^{1}$, the following equality holds
\begin{eqnarray}\label{efg}
    &&(e^\mathfrak{a}f\ast\widetilde{f})_0=(e^\mathfrak{a}f)_0(e^\mathfrak{a}\widetilde{f})_0,
\end{eqnarray}
where such a notation was used
\begin{eqnarray}\label{series}
    &&(e^\mathfrak{a}f)_0=\sum\limits_{n=0}^{\infty}\frac{1}{n!}
       \int\limits_{(\mathbb{R}^{3}\times\mathbb{R}^{3})^{n}}dx_{1}\ldots dx_{n}\,f_{n}(x_1,\ldots,x_n).
\end{eqnarray}


Now let us prove the equivalence of definition (\ref{ms}) of the reduced distribution
functions and their definition (\ref{FClusters}) within the framework of the dynamics
of correlations.

In terms of mapping (\ref{oper_d}) and notation (\ref{series}) the definition of reduced distribution
functions (\ref{ms}) is written as follows
\begin{eqnarray*}
    &&F_{s}(t,x_1,\ldots,x_s)=(e^\mathfrak{a}D(t))^{-1}_0(e^\mathfrak{a}\mathfrak{d}_{Y}D(t))_0.
\end{eqnarray*}
Using generalized cluster expansions (\ref{gcea}), and as a consequence of equalities
(\ref{d_gamma}) and (\ref{efg}), we find
\begin{eqnarray*}
    &&(e^\mathfrak{a}\mathfrak{d}_{Y}D(t))_0=
       (e^\mathfrak{a}\mathfrak{d}_{\{Y\}}{\mathbb E}\mathrm{xp}_{\ast}\,g(t))_0=\\
    &&(e^\mathfrak{a}\mathbb{E}\mathrm{xp}_{\ast}g(t)\ast\mathfrak{d}_{\{Y\}}g(t))_0
       =(e^\mathfrak{a}\mathbb{E}\mathrm{xp}_{\ast}g(t))_0(e^\mathfrak{a}\mathfrak{d}_{\{Y\}}g(t))_0.
\end{eqnarray*}
Taking into account that, according to the particular case $Y=\emptyset,$ of cluster
expansions (\ref{a8}), the equality holds
\begin{eqnarray*}
    &&(e^\mathfrak{a}\mathbb{E}\mathrm{xp}_{\ast}g(t))_0=(e^\mathfrak{a}D(t))_0,
\end{eqnarray*}
as a result, we establish the following representation for the reduced distribution functions
\begin{eqnarray*}
    &&F_{s}(t,x_1,\ldots,x_s)=(e^\mathfrak{a}\mathfrak{d}_{\{Y\}}g(t))_0.
\end{eqnarray*}
Therefore, in componentwise-form we obtain relation (\ref{FClusters}).

Since the correlation functions $g_{1+n}(t),\,n\geq0,$ are governed by the corresponding
Liouville hierarchy for the cluster of hard spheres and hard spheres, the reduced distribution
functions (\ref{FClusters}) are governed by the BBGKY hierarchy for hard spheres
\begin{eqnarray}\label{BBG}
    &&\frac{\partial}{\partial t}F(t)=e^{\mathfrak{a}}\mathcal{L}(\{\cdot\},\cdot\mid e^{-\mathfrak{a}}F(t)),
\end{eqnarray}
where the operator $\mathcal{L}(\{\cdot\},\cdot\mid f)$ is generator (\ref{LcL}) of the Liouville hierarchy
for a cluster of hard spheres and hard spheres. For a generator of this hierarchy of evolution equations
takes place the following representation:
\begin{eqnarray*}
    &&e^{\mathfrak{a}}\mathcal{L}(\{\cdot\},\cdot\mid e^{-\mathfrak{a}}F(t))=
      e^\mathfrak{a}\mathcal{L}^\ast e^{-\mathfrak{a}}F(t),
\end{eqnarray*}
where the operator $\mathcal{L}^\ast=\oplus_{n=0}^\infty\mathcal{L}_n^\ast$ is a direct sum of the Liouville
operators (\ref{Lstar}) and the operator $\mathfrak{a}$ is defined by formula (\ref{a}). Due to the fact that
pairwise collisions occur during the evolution, a generator of this hierarchy is reduced to the operator
of such a structure \cite{CGP97}
\begin{eqnarray*}
    &&e^\mathfrak{a}\mathcal{L}^\ast e^{-\mathfrak{a}}=\mathcal{L}^\ast + [\mathfrak{a},\mathcal{L}^\ast],
\end{eqnarray*}
where the bracket $[\cdot,\cdot]$ is the commutator of operators. For $t\geq0$ the collision part
$[\mathfrak{a},\mathcal{L}^\ast]$ of a generator of the BBGKY hierarchy for hard spheres has the following
explicit form
\begin{eqnarray*}
    &&\hskip-12mm \big([\mathfrak{a},\mathcal{L}^\ast]F(t)\big)_{s}(x_1,\ldots,x_s)=
      \sigma^{2}\sum\limits_{i=1}^s\int\limits_{\mathbb{R}^3\times\mathbb{S}_{+}^{2}}
      d p_{s+1}d\eta\,\langle\eta,(p_i-p_{s+1})\rangle \big(F_{s+1}(t,x_1,\ldots, \\
     &&q_i,p_i^{*},\ldots,x_s,q_i-\sigma\eta,p_{s+1}^{*})-
      F_{s+1}(t,x_1,\ldots,x_s,q_i+\sigma\eta,p_{s+1})\big),
\end{eqnarray*}
where the notations adopted above were used.

We note that the BBGKY hierarchy for hard spheres (\ref{BBG}) was mathematically justified in paper
\cite{PG90} (see also \cite{CGP97}).

In consequence of definition (\ref{FClusters}) and the cumulant structure of representation of a
solution (\ref{ghs}) for the Liouville hierarchy (\ref{Lh}), if initial state specified by the sequence
of reduced distribution functions $F(0)=(1,F_{1}^{0}(x_1),\ldots,$ $F_{n}^{0}(x_1,\ldots,x_n),\ldots)$,
then the evolution of all possible states, i.e. the sequence of the reduced distribution functions
$F_{s}(t),\,s\geq1$, is determined by the following series expansions \cite{GerRS}:
\begin{eqnarray}\label{RozvBBGKY}
   &&\hskip-12mm F_{s}(t,x_1,\ldots,x_s)=\sum\limits_{n=0}^{\infty}\frac{1}{n!}\,
       \int\limits_{(\mathbb{R}^{3}\times\mathbb{R}^{3})^{n}}dx_{s+1}\ldots dx_{s+n}\,
       \mathfrak{A}_{1+n}(t,\{1,\ldots,s\},\\
   &&\hskip+8mm  s+1,\ldots,{s+n})F_{s+n}^{0}(x_1,\ldots,x_{s+n}),\quad s\geq1,\nonumber
\end{eqnarray}
where the generating operator of these series
\begin{eqnarray}\label{cumulant1+n}
   &&\hskip-12mm \mathfrak{A}_{1+n}(t,\{1,\ldots,s\},s+1,\ldots,{s+n})=\\
   && \sum_{\mbox{\scriptsize$\begin{array}{c}\mathrm{P}:(\{x_1,\ldots,x_s\},\\
      x_{s+1},\ldots,x_{s+n})=\bigcup_i X_i\end{array}$}}
      (-1)^{|\mathrm{P}|-1}(|\mathrm{P}|-1)!\prod_{X_i\subset\mathrm{P}}
      S_{|\theta(X_i)|}(-t,\theta(X_i))\nonumber
\end{eqnarray}
is the $(1+n)th$-order cumulant (\ref{cumulantP}) of the groups of operators (\ref{Sspher})
and the notations adopted above was used.

We remark that the representation (\ref{RozvBBGKY}) is directly established for the initial
states satisfying the chaos condition due to the validity in this case of the representation
(\ref{gcph}) for the correlation functions of the hard-sphere cluster and of the hard spheres.

Consequently, as follows from the above, the cumulant structure of generating operators of
expansions for correlation functions (\ref{ghs}) or (\ref{rozvL-Nclusters}) induces the cumulant
structure (\ref{cumulant1+n}) of generating operators of series expansions for reduced distribution
functions (\ref{RozvBBGKY}) or in other words, the evolution of the state of a system of an infinite
number of hard spheres is governed by the dynamics of correlations on a microscopic scale.

Thus, we have established relation (\ref{FClusters}) between the reduced distribution functions
and correlation functions.

\subsection{Evolution of states described by the dynamics of correlations:\\ reduced correlation functions}
As is known, on a microscopic scale, the macroscopic characteristics of fluctuations of observables
are directly determined by means of the reduced correlation functions (marginal or $s$-particle
correlation functions \cite{Bog}, or cumulants of marginals \cite{TSRS20},\cite{DS}). Assuming
as a basis an alternative approach to the description of the evolution of states of a hard-sphere
system within the framework of correlation functions (\ref{ghs}), then the reduced correlation
functions are defined by means of a solution of the Cauchy problem of the Liouville hierarchy
(\ref{Lh}),(\ref{Lhi}) as follows:
\begin{eqnarray}\label{Gexpg}
   &&\hskip-12mm G_{s}(t,x_1,\ldots,x_s)\doteq \sum\limits_{n=0}^{\infty}\frac{1}{n!}\,
      \int\limits_{(\mathbb{R}^{3}\times\mathbb{R}^{3})^{n}}dx_{s+1}\ldots dx_{s+n}
      \,g_{s+n}(t,x_1,\ldots,x_{s+n}),\quad s\geq1,
\end{eqnarray}
where the generating function $g_{s+n}(t,x_1,\ldots,x_{s+n})$ is defined by expansion (\ref{ghs}),
or in terms of mapping (\ref{oper_d}) and notation (\ref{series}) this definition takes the form
\begin{eqnarray*}
   &&G_{s}(t,x_1,\ldots,x_s)=(e^\mathfrak{a}\mathfrak{d}_{Y}g(t))_0,
\end{eqnarray*}
or in terms of sequences of functions this expression has the form
\begin{eqnarray*}
   &&G(t)=e^\mathfrak{a}g(t).
\end{eqnarray*}
We emphasize that $nth$ term of expansions (\ref{Gexpg}) of the reduced correlation functions are
determined by the $(s+n)th$-particle correlation function (\ref{ghs}) in contrast with the expansions
of reduced distribution functions (\ref{FClusters}) which are determined by the $(1+n)th$-particle
correlation function of clusters of hard spheres (\ref{rozvL-Nclusters}).

Such a representation for reduced correlation functions (\ref{Gexpg}) can be derived as a result
of the fact that the reduced correlation functions are cumulants of reduced distribution functions
(\ref{FClusters}). Indeed, traditionally reduced correlation functions are introduced by means of
the cluster expansions of the reduced distribution functions similar to the cluster expansions of
the probability distribution functions (\ref{D_(g)N}) and on the set of allowed configurations
$\mathbb{R}^{3n}\setminus \mathbb{W}_n$ they have the form:
\begin{eqnarray}\label{FG}
   &&\hskip-8mm F_{s}(t,x_1,\ldots,x_s)=
      \sum\limits_{\mbox{\scriptsize$\begin{array}{c}\mathrm{P}:(x_1,\ldots,x_s)=\bigcup_{i}X_{i}\end{array}$}}
      \prod_{X_i\subset\mathrm{P}}G_{|X_i|}(t,X_i), \quad s\geq1,
\end{eqnarray}
where as above the symbol ${\sum\limits}_{\mathrm{P}:(x_1,\ldots,x_s)=\bigcup_{i} X_{i}}$ is the sum
over all possible partitions $\mathrm{P}$ of the set $(x_1,\ldots,x_s)$ into $|\mathrm{P}|$ nonempty
mutually disjoint subsets $X_i\subset(x_1,\ldots,x_s)$. As a consequence of this, the solution of
recurrence relations (\ref{FG}) are represented through reduced distribution functions as follows:
\begin{eqnarray}\label{gBigfromDFB}
   &&\hskip-8mm G_{s}(t,x_1,\ldots,x_s)=
   \sum\limits_{\mbox{\scriptsize $\begin{array}{c}\mathrm{P}:(x_1,\ldots,x_s)=\bigcup_{i}X_{i}\end{array}$}}
     (-1)^{|\mathrm{P}|-1}(|\mathrm{P}|-1)!\prod_{X_i\subset\mathrm{P}}F_{|X_i|}(t,X_i), \quad s\geq1.
\end{eqnarray}
Functions (\ref{gBigfromDFB}) are interpreted as the functions which describe the correlations of
hard-sphere states. The structure of expansions (\ref{gBigfromDFB}) is such that the reduced correlation
functions are cumulants (semi-invariants) of the reduced distribution functions (\ref{RozvBBGKY}).

Thus, taking into account representation (\ref{FClusters}) of the reduced distribution functions,
in consequence of the validity of relations (\ref{gClusters}) we derive representation (\ref{Gexpg})
of the reduced correlation functions through correlation functions
\begin{eqnarray*}
    &&\hskip-8mm G_{s}(t,x_1,\ldots,x_s)=
      \sum\limits_{\mathrm{P}:(x_1,\ldots,x_s)=\bigcup_{i}X_{i}}(-1)^{|\mathrm{P}|-1}(|\mathrm{P}|-1)!\,
      \prod_{X_i\subset\mathrm{P}}(e^\mathfrak{a}\mathfrak{d}_{\{X_i\}}g(t))=(e^\mathfrak{a}\mathfrak{d}_{Y}g(t))_0.
\end{eqnarray*}

Since the correlation functions $g_{s+n}(t),\,n\geq0,$ are governed by the Liouville hierarchy
for hard spheres (\ref{Lh}), the reduced correlation functions defined as (\ref{Gexpg}) are
governed by the hierarchy of nonlinear equations for hard spheres (the nonlinear BBGKY hierarchy):
\begin{eqnarray}\label{gBigfromDFBa}
   &&\hskip-12mm\frac{\partial}{\partial t}G_s(t,x_1,\ldots,x_s)=\mathcal{L}^{\ast}_{s}G_{s}(t,x_1,\ldots,x_s)+\\
   && \sum\limits_{\mathrm{P}:\,(x_1,\ldots,x_s)=X_{1}\bigcup X_2}\,\sum\limits_{i_{1}\in\widehat{X}_{1}}
      \sum\limits_{i_{2}\in \widehat{X}_{2}}\mathcal{L}_{\mathrm{int}}^{\ast}(i_{1},i_{2})
      G_{|X_{1}|}(t,X_{1})G_{|X_{2}|}(t,X_{2}))+\nonumber\\
   &&\int\limits_{\mathbb{R}^{3}\times\mathbb{R}^{3}}dx_{s+1}
      \big(\sum_{i=1}^{s}\mathcal{L}^{\ast}_{\mathrm{int}}(i,s+1)G_{s+1}(t,x_1,\ldots,x_{s+1})+\nonumber\\
   && \sum\limits_{\mathrm{P}:\,(x_1,\ldots,x_{s+1})=X_{1}\bigcup X_2}
      \sum_{i\in\widehat{X}_1;s+1\in \widehat{X}_2}
      \mathcal{L}^{\ast}_{\mathrm{int}}(i,s+1)G_{|X_1|}(t,X_1)G_{|X_2|}(t,X_2)\big), \nonumber\\ \nonumber\\
 \label{gBigfromDFBai}
   &&\hskip-12mmG_{s}(t,x_1,\ldots,x_s)\big|_{t=0}=G_{s}^{0}(x_1,\ldots,x_s), \quad s\geq1,
\end{eqnarray}
where the symbol
$\sum_{\mbox{\scriptsize $\begin{array}{c}\mathrm{P}:(x_1,\ldots,x_{s+1})=X_1\bigcup X_2,\end{array}$}}$
means the sum over all possible partitions of the set $(x_1,\ldots,x_{s+1})$ into two mutually disjoint
subsets $X_1$ and $X_2$, the sum over the index $i$ which takes values from the subset $\widehat{X}_1$
provided that the index $s+1$ belongs to the subset $\widehat{X}_2$ is denoted by
$\sum_{\mbox{\scriptsize $\begin{array}{c}i\in\widehat{X}_1;s+1\in\widehat{X}_2\end{array}$}}$ and
notations accepted in the Liouville hierarchy (\ref{Lh}) are used.

A generator of this hierarchy of nonlinear evolution equations has the following structure:
\begin{eqnarray*}\label{B}
    &&\frac{\partial}{\partial t}G(t)=e^{\mathfrak{a}}\mathcal{L}(\cdot\mid e^{-\mathfrak{a}}G(t)),
\end{eqnarray*}
where the operator $\mathcal{L}(\cdot\mid f)=\oplus_{n=0}^\infty\mathcal{L}(1,\ldots,n\mid f)$
is a direct sum of generators (\ref{LL}) of the Liouville hierarchy (\ref{Lh}).
Here are some component-wise examples of hierarchy (\ref{gBigfromDFBa}):
\begin{eqnarray*}
   &&\hskip-12mm\frac{\partial}{\partial t}G_1(t,x_1)=\mathcal{L}^{\ast}_{1}(1)G_{1}(t,x_1)+\\
   &&\int\limits_{\mathbb{R}^{3}\times\mathbb{R}^{3}}dx_{2}
      \mathcal{L}^{\ast}_{\mathrm{int}}(1,2)\big(G_{2}(t,x_1,x_{2})+
      G_{1}(t,x_1)G_{1}(t,x_2)\big),
\end{eqnarray*}
\begin{eqnarray*}
   &&\hskip-12mm\frac{\partial}{\partial t}G_2(t,x_1,x_2)=\mathcal{L}^{\ast}_{2}(1,2)G_{2}(t,x_1,x_2)+
        \mathcal{L}_{\mathrm{int}}^{\ast}(1,2)G_{1}(t,x_{1})G_{1}(t,x_{2})+\nonumber\\
   &&\int\limits_{\mathbb{R}^{3}\times\mathbb{R}^{3}}dx_{3}
     \Big(\sum_{i=1}^2\mathcal{L}^{\ast}_{\mathrm{int}}(i,3)\big(G_{3}(t,x_1,x_2,x_{3})+
     G_{2}(t,x_1,x_2)G_{1}(t,x_3)\big)+\nonumber\\
   && \mathcal{L}^{\ast}_{\mathrm{int}}(2,3)G_{2}(t,x_1,x_3)G_{1}(t,x_2)+
      \mathcal{L}^{\ast}_{\mathrm{int}}(1,3)G_{2}(t,x_2,x_3)G_{1}(t,x_1)\Big),\nonumber
\end{eqnarray*}
where it was used notations accepted above in definition (\ref{Lint}).

If $G(0)=(1,G_1^{0}(x_1),\ldots,G_s^{0}(x_1,\ldots,x_s),\ldots)$ is a sequence of reduced correlation
functions at initial instant, then by means of mappings (\ref{Sspherc}) the evolution of all possible
states, i.e. the sequence of the reduced correlation functions $G_{s}(t),\,s\geq1$, is determined by
the following series expansions:
\begin{eqnarray}\label{sss}
    &&\hskip-12mm G_{s}(t,x_1,\ldots,x_s)=\\
    &&\sum\limits_{n=0}^{\infty}\frac{1}{n!}
        \,\int\limits_{(\mathbb{R}^{3}\times\mathbb{R}^{3})^{n}}dx_{s+1}\ldots dx_{s+n}\,
        \,\mathfrak{A}_{1+n}(t;\{1,\ldots,s\},s+1,\ldots,s+n\mid G(0)), \quad s\geq1,\nonumber
\end{eqnarray}
where the generating operator $\mathfrak{A}_{1+n}(t;\{1,\ldots,s\},s+1,\ldots,s+n\mid G(0))$ of this
series is the $(1+n)th$-order cumulant of groups of nonlinear operators (\ref{ghs}):
\begin{eqnarray}\label{cc}
   &&\hskip-8mm\mathfrak{A}_{1+n}(t;\{1,\ldots,s\},s+1,\ldots,s+n\mid G(0))\doteq\\
   &&\sum\limits_{\mathrm{P}:\,(\{1,\ldots,s\},s+1,\ldots,s+n)=
      \bigcup_k X_k}(-1)^{|\mathrm{P}|-1}({|\mathrm{P}|-1})!
      \mathcal{G}(t;\theta(X_1)\mid\ldots \mathcal{G}(t;\theta(X_{|\mathrm{P}|})\mid G(0))\ldots), \nonumber\\
   &&\hskip-8mm   n\geq0,\nonumber
\end{eqnarray}
and where the composition of mappings (\ref{ghs}) of the corresponding noninteracting groups of
particles was denoted by
$\mathcal{G}(t;\theta(X_1)\mid \ldots\mathcal{G}(t;\theta(X_{|\mathrm{P}|})\mid G(0))\ldots)$,
for example,
\begin{eqnarray*}
    &&\hskip-5mm\mathcal{G}\big(t;1\mid\mathcal{G}(t;2\mid G(0))\big)=
        \mathfrak{A}_{1}(t,1)\mathfrak{A}_{1}(t,2)G^{0}_{2}(x_1,x_2),\\
    &&\hskip-5mm\mathcal{G}\big(t;1,2\mid\mathcal{G}(t;3\mid G(0))\big)=
        \mathfrak{A}_{1}(t,\{1,2\})\mathfrak{A}_{1}(t,3)G^{0}_{3}(x_1,x_2,x_3)+\\
    &&\hskip+5mm\mathfrak{A}_{2}(t,1,2)\mathfrak{A}_{1}(t,3)
       \big(G^{0}_{1}(x_1)G^{0}_{2}(x_2,x_3)+G^{0}_{1}(x_2)G^{0}_{2}(x_1,x_3)\big).
\end{eqnarray*}

We will adduce examples of expansions (\ref{cc}). The first order cumulant of the groups
of nonlinear operators (\ref{ghs}) is the group of these nonlinear operators
\begin{eqnarray*}
     &&\mathfrak{A}_{1}(t;\{1,\ldots,s\}\mid G(0))=\mathcal{G}(t;1,\ldots,s\mid G(0)).
\end{eqnarray*}
In case of $s=2$ the second order cumulant of nonlinear operators (\ref{ghs}) has the structure
\begin{eqnarray*}
     &&\hskip-5mm \mathfrak{A}_{1+1}(t;\{1,2\},3\mid G(0))=\mathcal{G}(t;1,2,3\mid G(0))-
       \mathcal{G}\big(t;1,2\mid\mathcal{G}(t;3\mid G(0))\big)=\\
     && \mathfrak{A}_{1+1}(t,\{1,2\},3)G^{0}_{3}(1,2,3)+\\
     && \big(\mathfrak{A}_{1+1}(t,\{1,2\},3)-
        \mathfrak{A}_{2}(t,2,3)\mathfrak{A}_{1}(t,1)\big)G^{0}_{1}(x_1)G^{0}_{2}(x_2,x_3)+\\
     &&\big(\mathfrak{A}_{1+1}(t,\{1,2\},3)-
        \mathfrak{A}_{2}(t,1,3)\mathfrak{A}_{1}(t,2)\big)G^{0}_{1}(x_2)G^{0}_{2}(x_1,x_3)+\\
     && \mathfrak{A}_{1+1}(t,\{1,2\},3)G^{0}_{1}(x_3)G^{0}_{2}(x_1,x_2)+
        \mathfrak{A}_{3}(t,1,2,3)G^{0}_{1}(x_1)G^{0}_{1}(x_2)G^{0}_{1}(x_3),
\end{eqnarray*}
where the operator
\begin{equation*}
    \mathfrak{A}_{3}(t,1,2,3)=\mathfrak{A}_{1+1}(t,\{1,2\},3)-
       \mathfrak{A}_{2}(t,2,3)\mathfrak{A}_{1}(t,1)-
       \mathfrak{A}_{2}(t,1,3)\mathfrak{A}_{1}(t,2)
\end{equation*}
is the third-order cumulant (\ref{cumcp}) of groups of operators (\ref{Sspher}) of a system
of hard spheres.

If $G(0)\in\oplus_{n=0}^{\infty}L^{1}_{n}$, then provided that
$\max_{n\geq1}\big\|G_n^{0}\big\|_{L^{1}_{n}}<(2e^{3})^{-1}$, for $t\in\mathbb{R}$ the sequence
of reduced correlation functions (\ref{sss}) is a unique solution of the Cauchy problem of the
nonlinear BBGKY hierarchy (\ref{gBigfromDFBa}),(\ref{gBigfromDFBai}) for hard spheres.

In the particular case of the initial state specified by the sequence of reduced correlation
functions $G^{(c)}=(0,G_1^{0},0,\ldots,0,\ldots)$ on the allowed configurations, that is, in
the absence of correlations between hard spheres at the initial moment of time \cite{CGP97},\cite{Sh},
according to definition (\ref{cc}) of the generating operators, reduced correlation functions
(\ref{sss}) are represented by the following series expansions:
\begin{eqnarray}\label{mcc}
   &&\hskip-12mm G_{s}(t,x_1,\ldots,x_s)=\\
   &&\sum\limits_{n=0}^{\infty}\frac{1}{n!}
     \,\int\limits_{(\mathbb{R}^{3}\times\mathbb{R}^{3})^{n}}dx_{s+1}\ldots dx_{s+n}\,
     \mathfrak{A}_{s+n}(t;1,\ldots,s+n)\prod_{i=1}^{s+n}G_1^{0}(x_i)
      \mathcal{X}_{\mathbb{R}^{3(s+n)}\setminus \mathbb{W}_{s+n}}, \quad s\geq1, \nonumber
\end{eqnarray}
where the generating operator $\mathfrak{A}_{s+n}(t)$ is the $(s+n)th$-order cumulant (\ref{cumcp})
of the groups of operators (\ref{Sspher}) of the Liouville equations (\ref{Liouville}).

We emphasize that in the absence of correlations of states of hard spheres on allowed configurations
at the initial moment of time, the generators of expansions into a series of reduced correlation
functions (\ref{mcc}) and reduced distribution functions (\ref{RozvBBGKY}) differ only in the order
of cumulants of groups of operators of hard spheres. Therefore, by means of such reduced distribution
functions or reduced correlation functions, the process of creating correlations in a system of hard
spheres is described.

We note that the reduced correlation functions give an equivalent approach to the description of the
evolution of states of many hard spheres along with the reduced distribution functions. Indeed, the
macroscopic characteristics of fluctuations of observables are directly determined by the reduced
correlation functions on the microscopic scale \cite{B49} for example, the functional of the dispersion 
of an additive-type observable, i.e. the sequence
$A^{(1)}=(0,a_{1}(x_1),\ldots,\sum_{i_{1}=1}^{n}a_1(x_{i_{1}}),\ldots)$, is represented by the formula
\begin{eqnarray*}
    &&\hskip-8mm \langle(A^{(1)}-\langle A^{(1)}\rangle)^2\rangle(t)=\\
    &&\hskip+7mm
      \int\limits_{\mathbb{R}^{3}\times\mathbb{R}^{3}}dx_{1}\,(a_1^2(x_1)-
      \langle A^{(1)}\rangle^2(t))G_{1}(t,x_1)+
      \int\limits_{(\mathbb{R}^{3}\times\mathbb{R}^{3})^2}dx_{1}dx_{2}\,a_{1}(x_1)a_{1}(x_2)G_{2}(t,x_1,x_2),
\end{eqnarray*}
where 
\begin{eqnarray*}
    &&\hskip-8mm \langle A^{(1)}\rangle(t)=\int_{\mathbb{R}^{3}\times\mathbb{R}^{3}}dx_{1}\,a_{1}(x_1)G_{1}(t,x_1)
\end{eqnarray*}
is the mean value functional of an additive-type observable.


\textcolor{blue!50!black}{\section{The description of correlations by means of kinetic equations}}

\subsection{The non-Markovian Enskog equation}
Now we will consider an approach to the description of the state evolution by means of
the state of a typical particle of a system of many hard spheres, or in other words,
foundations are overviewed of describing the evolution of a state by kinetic equations.

Let the initial state specified by a one-particle reduced correlation function, namely,
the initial state specified by a sequence of reduced correlation functions satisfying
a chaos property stated above, i.e. by the sequence $G^{(c)}=(G_0,G_1^{0},0,\ldots,0,\ldots)$
on the allowed configurations. We remark that such an assumption about initial states is
intrinsic in the contemporary kinetic theory of many-particle systems \cite{CGP97},\cite{CIP}.

Since the initial data $G^{(c)}$ is completely specified by a one-particle correlation
function, the Cauchy problem of the nonlinear BBGKY hierarchy (\ref{gBigfromDFBa}),(\ref{gBigfromDFBai})
for hard spheres is not a completely well-defined Cauchy problem, because the initial data
is not independent for every unknown function governed by the hierarchy of mentioned evolution
equations.
As a result, it becomes possible to reformulate such a Cauchy problem as a new Cauchy problem
for a one-particle correlation function with independent initial data and explicitly defined
functionals of the solution of this Cauchy problem.

We formulate such a restated Cauchy problem and the sequence of the suitable functionals.
The following statement is true.

In the case of the initial state $G^{(c)}$ specified by a one-particle correlation function
the evolution that described by means of the sequence $G(t)=\left(G_0,G_1(t),\ldots,G_s(t),\ldots\right)$
of reduced correlation functions (\ref{sss}), is also be described by the sequence
$G(t\mid G_{1}(t))=(G_0,G_1(t),\ldots,G_s(t\mid G_{1}(t)),\ldots)$ of the reduced (marginal) correlation
functionals: $G_s(t,x_1,\ldots,x_s\mid G_{1}(t)),\,s\geq2$, with respect to the one-particle correlation
function $G_1(t)$ governed by the non-Markovian Enskog kinetic equation.

Indeed, in the case under consideration the reduced correlation functionals $G_s(t\mid G_{1}(t)),\,s\geq2$,
are represented with respect to the one-particle correlation function (\ref{mcc}), i.e.
\begin{eqnarray}\label{ske}
   &&\hskip-8mm G_{1}(t,x_1)=\sum\limits_{n=0}^{\infty}\frac{1}{n!}\,
      \int\limits_{(\mathbb{R}^{3}\times\mathbb{R}^{3})^{n}}dx_{2}\ldots dx_{1+n}\,
      \mathfrak{A}_{1+n}(t,1,\ldots, n+1)\prod_{i=1}^{n+1}G_{1}^{0}(x_i)
      \mathcal{X}_{\mathbb{R}^{3(n+1)}\setminus \mathbb{W}_{n+1}},
\end{eqnarray}
by the following series expansions:
\begin{eqnarray}\label{f}
     && \hskip-8mm G_{s}\bigl(t,x_1,\ldots,x_s\mid G_{1}(t)\bigr)=\\
     && \sum_{n=0}^{\infty }\frac{1}{n!}\,
         \int\limits_{(\mathbb{R}^{3}\times\mathbb{R}^{3})^{n}}dx_{s+1}\ldots dx_{s+n}\,
         \mathfrak{V}_{s+n}\bigl(t,1,\ldots,s+n\bigr)\prod_{i=1}^{s+n}G_{1}(t,x_i),\quad s\geq2.\nonumber
\end{eqnarray}
The generating operator $\mathfrak{V}_{s+n}(t),\,n\geq0$, of the $(s+n)th$-order of this series
is determined by the following expansion:
\begin{eqnarray}\label{skrrc}
   &&\hskip-12mm\mathfrak{V}_{s+n}\bigl(t,1,\ldots,s,s+1,\ldots,s+n\bigr)= n!\,
     \sum_{k=0}^{n}\,(-1)^k\,\sum_{n_1=1}^{n}\ldots\\
   &&\hskip-10mm \sum_{n_k=1}^{n-n_1-\ldots-n_{k-1}}\frac{1}{(n-n_1-\ldots-n_k)!}
       \hat{\mathfrak{A}}_{s+n-n_1-\ldots-n_k}(t,\nonumber\\
   &&\hskip-10mm 1,\ldots,s+n-n_1-\ldots-n_k)\prod_{j=1}^k\,\sum\limits_{\mbox{\scriptsize$\begin{array}{c}
       \mathrm{D}_{j}:Z_j=\bigcup_{l_j}X_{l_j},\\
       |\mathrm{D}_{j}|\leq s+n-n_1-\dots-n_j\end{array}$}}\frac{1}{|\mathrm{D}_{j}|!}\times\nonumber\\
   &&\hskip-10mm
       \sum_{i_1\neq\ldots\neq i_{|\mathrm{D}_{j}|}=1}^{s+n-n_1-\ldots-n_j}\,
       \prod_{X_{l_j}\subset \mathrm{D}_{j}}\,\frac{1}{|X_{l_j}|!}
       \hat{\mathfrak{A}}_{1+|X_{l_j}|}(t,i_{l_j},X_{l_j}),\nonumber
\end{eqnarray}
where $\sum_{\mathrm{D}_{j}:Z_j=\bigcup_{l_j} X_{l_j}}$ is the sum over all possible dissections
of the linearly ordered set $Z_j\equiv(s+n-n_1-\ldots-n_j+1,\ldots,s+n-n_1-\ldots-n_{j-1})$ on no
more than $s+n-n_1-\ldots-n_j$ linearly ordered subsets, the $(s+n)th$-order scattering cumulant
is defined by the formula
\begin{eqnarray*}
    &&\hskip-8mm\hat{\mathfrak{A}}_{s+n}(t,1,\ldots,s+n)\doteq
      \mathfrak{A}_{s+n}(t,1,\ldots,s+n)\mathcal{X}_{\mathbb{R}^{3(s+n)}\setminus \mathbb{W}_{s+n}}
      \prod_{i=1}^{s+n}\mathfrak{A}_{1}^{-1}(t,i),
\end{eqnarray*}
and notations accepted above were used.

We adduce simplest examples of generating operators (\ref{skrrc}):
\begin{eqnarray*}
   &&\hskip-7mm\mathfrak{V}_{s}(t,1,\ldots,s)=
      \mathfrak{A}_{s}(t,1,\ldots,s)\mathcal{X}_{\mathbb{R}^{3s}\setminus \mathbb{W}_{s}}
      \prod_{i=1}^{s}\mathfrak{A}_{1}^{-1}(t,i),\\
   &&\hskip-7mm\mathfrak{V}_{s+1}(t,1,\ldots,s,s+1)=\mathfrak{A}_{s+1}(t,1,\ldots,s+1)
      \mathcal{X}_{\mathbb{R}^{3(s+1)}\setminus \mathbb{W}_{s+1}}
      \prod_{i=1}^{s+1}\mathfrak{A}_{1}^{-1}(t,i)-\\
   &&\mathfrak{A}_{s}(t,1,\ldots,s)\mathcal{X}_{\mathbb{R}^{3s}\setminus \mathbb{W}_{s}}
      \prod_{i=1}^{s}\mathfrak{A}_{1}^{-1}(t,i)\sum_{j=1}^s\mathfrak{A}_{2}(t,j,s+1)
      \mathcal{X}_{\mathbb{R}^{6}\setminus \mathbb{W}_{2}}
      \mathfrak{A}_{1}^{-1}(t,j)\mathfrak{A}_{1}^{-1}(t,s+1).
\end{eqnarray*}

A method of the construction of reduced correlation functionals (\ref{f}) is based on the
application of the so-called kinetic cluster expansions \cite{GG} to generating operators
(\ref{cumcp}) of series (\ref{mcc}).
If $\|G_{1}(t)\|_{L^{1}(\mathbb{R}^{3}\times\mathbb{R}^{3})}<e^{-(3s+2)}$, for arbitrary
$t\in\mathbb{R}$ series (\ref{f}) converges in the norm of the space $L^{1}_{s}$ \cite{GG}.

We note that in the case of initial state specified by a one-particle correlation function
the reduced correlation functionals (\ref{f}) describe all possible correlations generated
by the dynamics of many hard spheres in terms of a one-particle correlation function.

If initial data $G_{1}^{0}\in L^{1}_1$, then for arbitrary $t\in\mathbb{R}$ one-particle
correlation function (\ref{ske}) is a weak solution of the Cauchy problem of the non-Markovian
Enskog kinetic equation
\begin{eqnarray}\label{gkec}
   &&\hskip-12mm\frac{\partial}{\partial t}G_{1}(t,x_1)=\mathcal{L}^{\ast}(1)G_{1}(t,x_1)+
      \int\limits_{\mathbb{R}^{3}\times\mathbb{R}^{3}}dx_{2}\,
      \mathcal{L}_{\mathrm{int}}^{\ast}(1,2)G_{1}(t,x_1)G_{1}(t,x_2)+\\
   &&\hskip+9mm \int\limits_{\mathbb{R}^{3}\times\mathbb{R}^{3}}dx_{2}\,
      \mathcal{L}_{\mathrm{int}}^{\ast}(1,2)G_{2}\bigl(t,x_1,x_2\mid G_{1}(t)\bigr),\nonumber\\
   \nonumber\\
 \label{gkeci}
   &&\hskip-12mm G_{1}(t,x_1)\big|_{t=0}=G_{1}^{0}(x_1),
\end{eqnarray}
where the first part of the collision integral in equation (\ref{gkec}) has the Boltzmann--Enskog
structure, and the second part of the collision integral is determined in terms of the two-particle
correlation functional represented by series expansion (\ref{f}) and it describes all possible
correlations which are created by hard-sphere dynamics and by the propagation of initial correlations
related to the forbidden configurations.

By virtue of definitions (\ref{Lint}),(\ref{Lstar}) of the generator of the non-Markovian
Enskog equation (\ref{gkec}), for $t>0$ the kinetic equation gets such explicit form
\begin{eqnarray*}
   &&\hskip-7mm\frac{\partial}{\partial t}G_{1}(t,x_1)=
      -\langle p_1,\frac{\partial}{\partial q_1}\rangle G_{1}(t,x_1)+\\
   &&\sigma^2\int\limits_{\mathbb{R}^3\times\mathbb{S}_{+}^{2}}d p_{2}d\eta\,\langle\eta,(p_{1}-p_{2})\rangle
        \big(G_{1}(t,p_{1}^\ast,q_1)G_{1}(t,p_{2}^\ast,q_{1}-\sigma\eta,)-G_{1}(t,x_1)G_{1}(t,p_2,q_1+\sigma\eta)\big)+\\
   &&\sigma^2\int\limits_{\mathbb{R}^3\times\mathbb{S}_{+}^{2}}d p_{2}d\eta\,\langle\eta,(p_{1}-p_{2})\rangle
       \big(G_{2}\bigl(t,p_{1}^\ast,q_1,p_{2}^\ast,q_{1}-\sigma\eta\mid G_{1}(t)\bigr)-
       G_{2}\bigl(t,x_1,p_2,q_1+\sigma\eta\mid G_{1}(t)\bigr)\big).
\end{eqnarray*}

In the paper \cite{GG}, similar statements were proved for the evolution of the state
of a hard-sphere system described in terms of reduced distribution functions governed
by the BBGKY hierarchy. We emphasize that the $nth$ term of expansions (\ref{f}) of the
reduced correlation functionals are determined by the $(s+n)th$-order generating operator
(\ref{skrrc}) in contradistinction to the expansions of reduced distribution functionals
of the state constructed in \cite{GG} which are determined by the $(1+n)th$-order generating
operator (\ref{skrrc}).

Thus, for the initial state specified by a one-particle correlation function all possible
states of a system of hard spheres can be described without any approximations within the
framework of a one-particle correlation function governed by the non-Markovian kinetic equation
(\ref{gkec}), and a sequence of explicitly defined functionals (\ref{f}) of its solution (\ref{ske}).


\subsection{On the low-density approximation of hard-sphere correlations}
The conventional philosophy of the description of kinetic evolution consists in the following.
If the initial state is specified by a one-particle distribution function, then the evolution
of the state can be effectively described in a suitable scaling limit \cite{Bog},\cite{G58}
by means of a one-particle distribution function governed by the nonlinear kinetic equation.

In the last decade, the Boltzmann--Grad limit (low-density scaling limit) \cite{G58},\cite{L75}
of the reduced distribution functions constructed by means of the theory of perturbations was
rigorously discussed in several papers, see for example \cite{TSRS20},\cite{PS16},\cite{GG21}
and references therein.
In this subsection, we consider a scheme of constructing the scaling asymptotic behavior of
reduced correlation functions (\ref{sss}) in the Boltzmann--Grad limit for initial data satisfying
a chaos property stated above, namely, for the sequence of reduced correlation functions (\ref{mcc}).

Let for $t\geq0$ the operator $\mathcal{L}^{\ast}_{\mathrm{int}}$ in the dimensionless form of the
hierarchy of evolution nonlinear equations (\ref{gBigfromDFBa}) be scaled in such a way that
\begin{eqnarray*}
     &&\hskip-7mm \mathcal{L}_{\mathrm{int}}^\ast(j_{1},j_{2})f_{n}=
        \epsilon^2\int\limits_{\mathbb{S}_{+}^2}d\eta\langle\eta,(p_{j_{1}}-p_{j_{2}})\rangle
        \big(f_n(x_1,\ldots,p_{j_{1}}^*,q_{j_{1}},\ldots,\\
     &&p_{j_{2}}^*,q_{j_{2}},\ldots,x_n)\delta(q_{j_{1}}-q_{j_{2}}+\epsilon\eta)-
        f_n(x_1,\ldots,x_n)\delta(q_{j_{1}}-q_{j_{2}}-\epsilon\eta)\big),\nonumber
\end{eqnarray*}
where $\epsilon>0$ is a scaling parameter (the ratio of the diameter $\sigma>0$ to the mean
free path of hard spheres) and the notations similar to definition (\ref{Lint}) are used.
We will consider initial states of a hard-sphere system specified by the scaled one-particle
correlation function $G_{1}^{0,\epsilon},$ such that:
$|G_{1}^{0,\epsilon}(x_1)|\leq ce^{\textstyle-\frac{\beta}{2}{p^{2}_1}},$
where $\beta>0$ is a parameter and $c<\infty$ is some constant. We emphasize that the states of
a system of infinitely many hard spheres are described by sequences of functions bounded with
respect to the configuration variables \cite{CGP97} as is assumed above. Regarding the technical
details of the scheme presented below for constructing the Boltzmann--Grad asymptotics of reduced
correlation functions (\ref{mcc}), we refer to our article \cite{PG90}.

Let us assume that the Boltzmann--Grad limit of the initial one-particle correlation function
$G_{1}^{0,\epsilon}$ exists in the sense of the weak convergence \cite{PG90}
\begin{eqnarray}\label{asic1}
   &&\mathrm{w-}\lim_{\epsilon\rightarrow 0}\big(\epsilon^{2}G_{1}^{0,\epsilon}(x_1)-f_{1}^0(x_1)\big)=0.
\end{eqnarray}
If equality (\ref{asic1}) holds for the initial one-particle correlation function, then for series
expansion (\ref{mcc}) the following equality is true
\begin{eqnarray*}
   &&\mathrm{w-}\lim\limits_{\epsilon\rightarrow 0}\big(\epsilon^{2}G_{1}(t,x_1)-f_{1}(t,x_1)\big)=0,
\end{eqnarray*}
where for some finite time interval the limit one-particle correlation function
$f_1(t,x_1)$ is represented by the series expansion
\begin{eqnarray}\label{1mco}
   &&\hskip-8mm f_{1}(t,x_1)=\sum\limits_{n=0}^{\infty}\int\limits_0^tdt_{1}\ldots
      \int\limits_0^{t_{n-1}}dt_{n}\,\int\limits_{(\mathbb{R}^{3}\times\mathbb{R}^{3})^{n}}dx_{2}\ldots dx_{1+n}\,
      S_{1}(t_{1}-t,1)\times \\
   &&\mathcal{L}^{\ast,0}_{\mathrm{int}}(1,2)\prod\limits_{j_1=1}^{2}S_{1}(t_{2}-t_{1},j_1)\ldots
      \prod\limits_{i_{n}=1}^{n}S_{1}(t_{n}-t_{n-1},i_{n})\times \nonumber\\
   &&\sum\limits_{k_{n}=1}^{n}\mathcal{L}^{\ast,0}_{\mathrm{int}}(k_{n},n+1)
      \prod\limits_{j_n=1}^{n+1}S_{1}(-t_{n},j_n)\prod\limits_{i=1}^{n+1}f_1^{0}(x_i).\nonumber
\end{eqnarray}
In this series expansion for $t\geq0$ the operator $\mathcal{L}_{\mathrm{int}}^{\ast,0}(j_{1},j_{2})$
is defined by the formula
\begin{eqnarray*}
     &&\hskip-7mm \mathcal{L}_{\mathrm{int}}^{\ast,0}(j_{1},j_{2})f_{n}
        \doteq\int\limits_{\mathbb{S}_{+}^2}d\eta\langle\eta,(p_{j_{1}}-p_{j_{2}})\rangle
        \big(f_n(x_1,\ldots,p_{j_{1}}^*,q_{j_{1}},\ldots,\\
     &&p_{j_{2}}^*,q_{j_{2}},\ldots,x_n)-f_n(x_1,\ldots,x_n)\big)\delta(q_{j_{1}}-q_{j_{2}}),\nonumber
\end{eqnarray*}
where notations adopted in formula (\ref{Lint}) are used.

The function $f_{1}(t)$ represented by series (\ref{1mco}) is a weak solution of the Cauchy problem
of the Boltzmann kinetic equation for hard spheres
\begin{eqnarray}\label{Bc}
   &&\hskip-7mm\frac{\partial}{\partial t}f_{1}(t,x_1)=
      -\langle p_1,\frac{\partial}{\partial q_1}\rangle f_{1}(t,x_1)+\\
   &&\hskip+5mm\int_{\mathbb{R}^3\times\mathbb{S}^2_+}d p_2\,d\eta
       \,\langle\eta,(p_1-p_2)\rangle(f_1(t,p_1^{*},q_1)f_1(t,p_2^{*},q_1)-f_1(t,p_1,q_1)f_1(t,p_2,q_1)), \nonumber\\ \nonumber\\
\label{Bci}
   &&\hskip-7mmf_{1}(t,x_1)\big|_{t=0}=f_{1}^0(x_1).
\end{eqnarray}

The property of propagation of initial chaos in the of low-density limit is a consequence of the following
equalities for reduced correlation functions (\ref{mcc}):
\begin{eqnarray}\label{dchaos}
   &&\mathrm{w-}\lim\limits_{\epsilon\rightarrow 0}\epsilon^{2s}G_{s}\bigl(t,x_1,\ldots,x_s\bigr)=0, \quad s\geq2.
\end{eqnarray}


The proof of these statements (\ref{1mco}) and (\ref{dchaos}) is based on the validity of the limit
theorems for cumulants of asymptotically perturbed groups of operators (\ref{Sspher}) and the
explicit structure of the generating operators of series expansions of reduced correlation functions
(\ref{mcc}).

Indeed, if $|f_{s}|\leq ce^{\textstyle-\frac{\beta}{2}{\sum^s_{i=1} p^{2}_i}}$, then for arbitrary
finite time interval for asymptotically perturbed first-order cumulant (\ref{cumcp}) of the groups
of operators (\ref{Sspher}) the following equality takes place \cite{CGP97},\cite{PG90}
\begin{eqnarray*}
    &&\mathrm{w-}\lim\limits_{\epsilon\rightarrow 0}\big(S_{s}(-t,1,\ldots,s)f_{s}-
        \prod\limits_{j=1}^{s}S_{1}(-t,j)f_{s}\big)=0.
\end{eqnarray*}
Therefore, for the $(s+n)th$-order cumulant of asymptotically perturbed groups of operators (\ref{Sspher})
the equalities are true:
\begin{eqnarray*}
   &&\mathrm{w-}\lim\limits_{\epsilon\rightarrow 0}\frac{1}{\epsilon^{2n}}\,
     \mathfrak{A}_{s+n}(t,1,\ldots,s+n)f_{s+n}=0, \quad s\geq2.
\end{eqnarray*}

In conclusion, we note that in the articles \cite{GG18},\cite{GG21} some other approaches to the derivation
of kinetic equations were developed, including the kinetic equations for the states with correlations at the
initial instant. In particular, a new method of the description of the kinetic evolution of a system with
hard-sphere collisions has been suggested in the paper \cite{GG18}. Formalism is based on the description
of evolution within the framework of observables which are governed by the dual Boltzmann hierarchy in the
low-density limit.


\textcolor{blue!50!black}{\section{Conclusion}}

This article dealt with the mathematical problems of the description of the evolution of many hard
spheres based on various ways of describing their state, namely by means of functions describing
the propagation of correlations. One of these approaches allows one to describe the evolution of
both a finite and an infinite average number of hard spheres using reduced distribution functions
(\ref{RozvBBGKY}) or reduced correlation functions (\ref{sss}), which are determined by the dynamics
of correlations (\ref{ghs}) of a hard-sphere system.

We note the importance of the description of the processes of the creation and propagation of
correlations \cite{P62}, in particular, it is related to the problem of the description of the
memory effects in many-particle systems with collision dynamics.

It was established that the notion of cumulants (\ref{cumulantP}) of the groups of operators
(\ref{Sspher}) underlies non-perturbative expansions of solutions for the fundamental evolution
equations describing the state evolution of a hard-sphere system, namely, of the Liouville
hierarchy (\ref{Lh}) for correlation functions, of the BBGKY hierarchy for reduced distribution
functions and of the nonlinear BBGKY hierarchy (\ref{gBigfromDFBa}) for reduced correlation
functions, as well as it underlies the kinetic description of infinitely many hard spheres (\ref{f}).
We remark that for quantum many-particle systems the concept of cumulants of groups of operators
is considered in the papers \cite{GerS}-\cite{G12}.

We emphasize that the structure of expansions for correlation functions (\ref{ghs}), in which
the generating operators are the cumulants of the corresponding order (\ref{cumulantP}) of the
groups of operators (\ref{Sspher}) of hard spheres, induces the cumulant structure of series
expansions for reduced distribution functions (\ref{RozvBBGKY}), reduced correlation functions
(\ref{sss}) and reduced correlation functionals (\ref{f}). Thus, in fact, the dynamics of
systems of infinitely many hard spheres is generated by the dynamics of correlations.

Above, one more approach to the description of the state evolution of a system of many hard
spheres in terms of the state evolution of a typical particle was also studied. In other words,
the origin of the collective behavior of a hard-sphere system on a microscopic scale was described
by means of a one-particle correlation function that is determined by the non-Markovian Enskog
kinetic equation (\ref{gkec}). As already mentioned, one of the advantages of such an approach to
the derivation of kinetic equations from underlying collisional dynamics consists of an opportunity
to construct the kinetic equations with initial correlations, which makes it possible to describe
the propagation of initial correlations in the Boltzmann--Grad limit \cite{GG21}. Another advantage
of this approach is related to the rigorous derivation of the Boltzmann equation with higher-order
corrections to the main term of the Boltzmann--Grad asymptotics of collisional dynamics.


\bigskip
\textbf{Acknowledgements.} \,\,\,{\large\textcolor{blue!55!black}{Glory to Ukra\"{\i}ne!}}

\bigskip

\addcontentsline{toc}{section}{{References}}
\renewcommand{\refname}{\textcolor{blue!50!black}{References}}
\small{

}
\end{document}